\documentclass[conference]{IEEEtran}
\usepackage{graphicx}
\usepackage{amssymb}
\usepackage{pifont}
\usepackage[all=normal, title=normal, paragraphs=tight, wordspacing=normal]{savetrees}
\usepackage{comment}
\usepackage{subfig}
\usepackage{algorithmic}
\usepackage{algorithm}
\usepackage{framed}
\usepackage{balance}

\usepackage{amsmath,amsfonts,bm}

\def\eqref#1{equation~\ref{#1}}

\def\1{\bm{1}}

\def\va{{\bm{a}}}
\def\vb{{\bm{b}}}

\def\vx{{\bm{x}}}

\def\mW{{\bm{W}}}
\def\mX{{\bm{X}}}
\def\mY{{\bm{Y}}}

\DeclareMathAlphabet{\mathsfit}{\encodingdefault}{\sfdefault}{m}{sl}
\SetMathAlphabet{\mathsfit}{bold}{\encodingdefault}{\sfdefault}{bx}{n}

\newcommand{\sigmoid}{\sigma}

\usepackage{hyperref}
\usepackage{cleveref}
\usepackage{url}
\usepackage{multirow}
\usepackage{booktabs}
\title{Dynamic program analysis with \\ Neural Networks}
\usepackage{amsmath}
\usepackage{listings}
\usepackage{graphicx}
\usepackage{amsmath}
\usepackage{xspace}
\usepackage{color}
\usepackage{longfbox}

\newcommand{\rom}[1]{\uppercase\expandafter{\romannumeral #1\relax}}

\newcommand{\eg}{\hbox{\emph{e.g.,}}\xspace}
\newcommand{\ie}{\hbox{\emph{i.e.}}\xspace}

\newcommand{\wrt}{\hbox{\emph{w.r.t.}}\xspace}

\newcommand{\dta}{\textsc{dta}\xspace}
\newcommand{\tool}{\textsc{Neutaint}\xspace}
\newcommand{\nn}{\textsc{NN}\xspace}

\definecolor{javared}{rgb}{0.6,0,0} 
\definecolor{javagreen}{rgb}{0.25,0.5,0.35} 
\definecolor{javapurple}{rgb}{0.5,0,0.35} 
\definecolor{javadocblue}{rgb}{0.25,0.35,0.75} 

\lstdefinestyle{customc}{
  belowcaptionskip=\baselineskip,
  breaklines=true,
  xleftmargin=\parindent,
  language=java,
  showstringspaces=false,
  basicstyle=\footnotesize\ttfamily,
  keywordstyle=\bfseries\color{javapurple},
  commentstyle=\itshape\blue,
 numbers=left,
 numbersep=.1cm, 
 linewidth=\columnwidth,
 xleftmargin=2.2em,
 framexleftmargin=1.1em,
 frame=single,
 float=H, 
 aboveskip=\baselineskip
}

\lstdefinestyle{CStyle}{
    backgroundcolor=\color{backgroundColour},   
    commentstyle=\color{mGreen},
    keywordstyle=\color{magenta},
    numberstyle=\tiny\color{mGray},
    stringstyle=\color{mPurple},
    basicstyle=\footnotesize,
    breakatwhitespace=false,         
    breaklines=true,                 
    captionpos=b,                    
    keepspaces=true,                 
    numbers=left,                    
    numbersep=5pt,                  
    showspaces=false,                
    showstringspaces=false,
    showtabs=false,                  
    tabsize=2,
    language=C
}

\definecolor{codegreen}{rgb}{0,0.6,0}
\definecolor{codegray}{rgb}{0.5,0.5,0.5}
\definecolor{codepurple}{rgb}{0.58,0,0.82}
\definecolor{backcolour}{rgb}{0.95,0.95,0.92}

\lstdefinestyle{braystyle}{
  commentstyle=\Red,
  keywordstyle=\blue,
  numberstyle=\tiny\color{codegray},
  stringstyle=\color{codepurple},
  basicstyle=\scriptsize,
  breakatwhitespace=false,         
  breaklines=true,                 
  captionpos=b,                    
  keepspaces=true,                 
  numbers=left,                    
  numbersep=4pt,                  
  showspaces=false,                
  showstringspaces=false,
  showtabs=false,                  
  tabsize=2,
  belowskip=-3
  \baselineskip,
  frame=None,
  language=C,
  float=H
}

\lstdefinestyle{braystyle1}{
  commentstyle=\Red,
  keywordstyle=\blue,
  numberstyle=\tiny\color{codegray},
  stringstyle=\color{codepurple},
  basicstyle=\scriptsize,
  breakatwhitespace=false,         
  breaklines=true,                 
  captionpos=b,                    
  keepspaces=true,                 
  numbers=left,                    
  numbersep=4pt,                  
  showspaces=false,                
  showstringspaces=false,
  showtabs=false,                  
  tabsize=2,
  belowskip=-3
  \baselineskip,
  frame=single,
  language=C,
  float=H
}

\newcommand*{\blue}[1]{\textcolor{blue}{#1}}

\newcommand{\authnote}[2]{{\bf \textcolor{blue}{#1}: \em \textcolor{red}{#2}}}

\newcommand{\dongdong}[1]{\authnote{Dongdong}{#1}}

\newcommand{\yizheng}[1]{\authnote{Yizheng}{#1}}
\renewcommand{\baselinestretch}{0.99}

\usepackage{float}
\usepackage{caption}

\newcommand\Red[1]{\textcolor[rgb]{1.00,0.00,0.00}{\textbf{#1}}}

\begin{document}
\title{Neutaint: Efficient Dynamic Taint Analysis\\ with Neural Networks} 

\author{\IEEEauthorblockN{Dongdong She, Yizheng Chen, Abhishek Shah, Baishakhi Ray and 
Suman Jana}
\IEEEauthorblockA{Columbia University\\
\
}
}

\maketitle
\thispagestyle{plain}

\pagestyle{plain}

\begin{abstract}
Dynamic taint analysis (DTA) is widely used by various applications to track information flow during runtime execution. Existing DTA techniques use rule-based taint-propagation, which is neither 
accurate (\ie, high false positive rate) nor efficient (\ie, large runtime overhead). It is hard to specify taint rules for each operation while covering all corner cases correctly. Moreover, the overtaint and undertaint errors can accumulate during the propagation of taint information across multiple operations. Finally, rule-based propagation requires each operation to be inspected before applying the appropriate rules resulting in prohibitive performance overhead on large real-world applications.  

In this work, we propose \tool, a novel end-to-end approach to track information flow using neural program embeddings. The neural program embeddings model the target's programs computations taking place between taint sources and sinks, which automatically learns the information flow by observing a diverse set of execution traces. To perform lightweight and precise information flow analysis, we utilize saliency maps to reason about most influential sources for different sinks. \tool constructs two saliency maps, a popular machine learning approach to influence analysis, to summarize both coarse-grained and fine-grained information flow in the neural program embeddings.

We compare \tool with 3 state-of-the-art dynamic taint analysis tools. 
The evaluation results show that \tool can achieve 68\% accuracy, on average, which is 10\% improvement while reducing 40$\times$ runtime overhead over the second-best taint tool Libdft on 6 real world programs. 
\tool also achieves 61\% more edge coverage when used for taint-guided fuzzing indicating the effectiveness of the identified influential bytes. 
We also evaluate \tool's ability to detect real world software attacks. The results show that \tool can successfully detect different types of vulnerabilities including buffer/heap/integer overflows, division by zero, etc. Lastly, \tool can detect $98.7\%$ of total flows, the highest among all taint analysis tools.

\end{abstract}

\section{Introduction}

Dynamic Taint Analysis (DTA)~\cite{Newsome05dynamictaint} is a well-known technique to track information flow between source and sink variables during a program's execution. It has been used in many security-relevant applications including guided fuzzing, automatic vulnerability discovery, run-time policy enforcement, information leak detection and malware behavior analysis.~\cite{dytan}~\cite{tainted_fuzz}~\cite{Kang2011DTADT}~\cite{lekies201325}~\cite{ melicher2018riding}~\cite{Newsome05dynamictaint}~\cite{Shankar}~\cite{Tripp_2009}~\cite{Vogt2007CrossSS}~\cite{Wang_2010}~\cite{Yin07panorama}. Most, if not all, applications of DTA require high accuracy and low run-time overhead. Unfortunately, existing DTA techniques suffer both from high false positive/negative rates and incur prohibitive performance overhead especially for large real-world programs~\cite{taintinduce}.


All existing DTA techniques \emph{propagate} taint labels from the taint source to the sinks during the target program's execution based on a set of rules for every executed statement. 
The final taint results are computed by propagating and composing the individual per-statement taint rules together. 
Essentially, the final output indicates whether a taint source influences a sink. 

Unfortunately, this rule-based propagation approach has three fundamental limitations: 
(i) {\em Specifying accurate propagation rules}: 
 Even for seemingly simple operations, accurately specifying propagation rules is often hard as there can be many different cases to consider. For instance, the correct propagation rule for \texttt{s = x*c} might vary based on different values of taint labels of \texttt{x} and constant \texttt{c}\textemdash if \texttt{c} is always 0, \texttt{s} is not influenced by \texttt{x}. Similarly, if \texttt{c} is very large and \texttt{x} is small, the influence of x on the output might be negligibly small. It is extremely difficult to enumerate all such possibilities exhaustively.
(ii) {\em Accumulating errors}: 
Even if taint propagation rules for each operation are accurate, their composition across multiple operations can introduce large errors. For example, consider two operations \texttt{s = a + b; t = s - b}, where the rule-based propagation will conclude that both \texttt{s} and \texttt{t} are influenced by \texttt{b}. Although this is the correct analysis for each operation individually, \texttt{t} is not affected by \texttt{b}.  
(iii) {\em Large run-time overhead}: The rule-based propagation introduces prohibitive run-time overhead as each operation has to be examined to decide which rules to apply. 

In this paper, we propose a novel technique, \tool that automatically learns the information flow, \ie, taint, 
in a program by modeling its source-sink behaviors with neural program embeddings and gradient analysis. 
Neural program embeddings are essentially neural networks that learn to predict program behaviors from different representations of a program (e.g., graph representation, input-output pairs)~\cite{neural_repair, pmlr-v37-piech15, neuzz, shen2018neuro}. Such embeddings have shown promise in various tasks including fuzzing, program repair, program synthesis, vulnerability localization and binary similarity detection~\cite{neural_repair, Gupta2017DeepFixFC,BhatiaS16,Pu2016, neuzz}. Leveraging the chain rule of calculus, gradient analysis is a more precise technique using automatic gradient computation to accurately track the influence of sources over sinks in programs. Our \tool first learns a neural program embedding of a program's run-time behaviors and then performs gradient analysis for light-weight and accurate end-to-end information flow tracking.

\tool addresses the aforementioned limitations of rule-based taint tracking. First, while rule-based DTA applies the propagation rules based on the program statements executed along a single flow, neural networks can generalize and infer new flows based on the past program behaviors. This allows us to more accurately model different degrees of influence from different taint sources. Second, since neural networks are continuous, the gradient computation provides an efficient and precise mathematical way of deriving a source's influence on a particular sink, thus avoiding the need to manually specify propagation rules. This also minimizes the composition errors that plague existing rule-based approaches and significantly improves the accuracy of taint tracking by cutting down the false positive/negative rates. Lastly, the neural program embeddings can be trained using program traces generated offline by adding light-weight instrumentation and executing the target program with multiple inputs. Once trained, the neural program embeddings can be used to perform taint analysis without even examining the computations performed by the target program in an highly efficient manner compared to rule-based propagation.

\begin{figure*}[!t]
\begin{tabular}{rl}
\begin{minipage}[t]{.15\linewidth}
\vspace{0.6cm}
\lstset{basicstyle=\small\ttfamily, breaklines=true, xleftmargin=-0.000cm}
\begin{lstlisting}[language={C}, captionpos=b]
 x = input();
 a = x[0];
 b = x[1];
 c = a*a + b;
 z = c - b;
 print z
\end{lstlisting} 

\vspace{1.55cm}\hspace{0.1cm}
\includegraphics[width=0.8\linewidth]{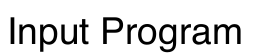}
\end{minipage}
&
\begin{minipage}[t]{0.82\linewidth}
\subfloat{\includegraphics[width=\linewidth]{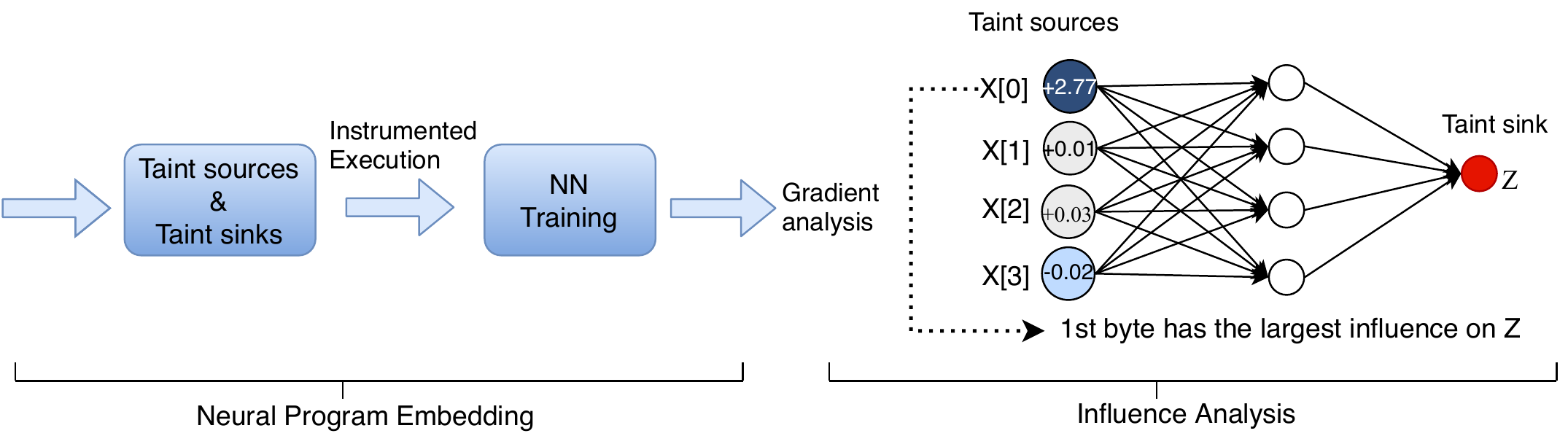}}
\end{minipage}
\\
\end{tabular}
\vspace{0.4cm}
\caption{Simple code snippet demonstrating the workflow of \tool. \tool uses light-weight instrumentation to collect a diverse set of sources and sinks from the input program. Then, we train neural program embeddings and use gradient-based analysis to infer information flow for the programs.}
\label{fig:workflow}

\end{figure*}


Specifically, we train dynamic neural program embeddings using observed execution paths between sources and sinks. Once we have observed one path reaching the sink from the source during program execution, we generate a lot of other program paths in a cheap way (e.g., mutating the input) to obtain the training data. 
Note that any form of DTA (rule-based or learning-based) requires an input triggering a program execution path from the source to the sink to begin its analysis. Our key advantage is that we can accurately infer new information flows without inspecting each executed statement. Even with simple training data, we can track more flows than three state-of-the-art DTA tools: Libdift, Triton, and DFSan. We can potentially further improve the training data quality by using techniques such as symbolic execution. We present detailed quantitative results showing the number of detected and missed flows by \tool in~\Cref{subsec:infoloss}.

After training the neural program, we use a gradient-based attribution method on the trained network to create saliency maps that accurately measure the flow of information from \nn inputs (\ie, taint sources) to \nn outputs (\ie, taint sinks). 
Depending on the application domain, \tool supports two types of saliency maps to track information flow: (i) a coarse-grained map that aggregates the influence of all sources over all sinks. This map contains the information flow summarized by all executed inputs over all taken paths. (ii) a fine-grained map containing separate influence analysis for each source-sink pair. Such information is useful for tasks like zero-day attack detection. 
A common use case for such analysis is taint-guided fuzzing where the sources are the input bytes and the sinks are the variables used in program branches. Input bytes with high saliency values have large influences on the output variables in program branches. Mutating these bytes can maximally trigger the execution of a diverse set of program branches. The number of mutations spent on each byte can be adjusted according to its corresponding aggregate influence on all program branches, \ie, the higher the influence, the more mutations should be tried on the corresponding byte.

We evaluate \tool against 3 state-of-the-art dynamic taint analysis tools: Libdft, Triton, and DFSan. We train neural network models representing two sets of real-world programs. For the first set of programs, \tool can successfully find the information flow from source to sink in known CVEs. On the second set of programs, we compare \tool against other tools in regards to accuracy, overhead, and effectiveness when applied to fuzzing (an important security application). We utilize the parsing logic of programs to build ground truth of hot bytes, \ie, file format bytes (influential taint sources) that trigger different program behaviors (taint sinks at branching conditions). The evaluation results show that \tool achieves on average 10\% higher taint accuracy than the second-best DTA tool. To compare the runtime overhead, we measure the total amount of time needed to process all the inputs in the training dataset, and our \tool is almost 40$\times$ more efficient than the second-fastest tool, Libdft. We then validate the taint information obtained from all tools through taint-guided fuzzing. We feed the hot bytes produced by the four different tools into a common fuzzer backend that supports the same mutation algorithm in which our \tool achieves 61\% more edge coverage on all the real-world programs in 24 hours. 

Our main contributions are as follows:

\begin{itemize}
\item
We propose a novel information flow tracking technique based on neural program embedding and gradient analysis.
  
\item We design and implement our technique as part of \tool and evaluate it against 3 state-of-the-art DTA tools. The evaluation shows that \tool can achieve on average 10\% higher taint accuracy than the second-best tool while taking 40$\times$ less analysis time.

\item
We further validate the taint information obtained from 4 different tools by using a real world taint application, taint-guided fuzzing. The results show that \tool achieves $61$\% more edge coverage than that of the second-best DTA tool.

\item
We analyze and identify the key factors that allow \tool to outperform traditional DTA tools. In addition, we present quantitative results showing \tool's ability to infer new information flows and discuss different ways to further improve the training data quality.

\end{itemize}

\section{Background}
\label{sec:back}

This section first gives a brief overview of dynamic taint analysis. Then, we introduce existing work in program embeddings, among which dynamic program embeddings can be used to capture the runtime program behavior. Lastly, we discuss saliency maps for neural networks, which can be used to conduct information flow analysis for dynamic program embeddings.

\medskip
\noindent
\textbf{Dynamic Taint Analysis (\dta).}
A dynamic taint analysis pre-defines taint sources (\eg untrusted file, network, etc.) and as a program executes tracks the effect of them on program state such as internal variables. In most cases, \dta wants to determine whether the taint sources affect some predefined target locations, commonly known as taint {\em sinks}. Depending on the specific application, taint sources and sinks vary.

For many security applications, user inputs are often used as taint sources~\cite{Newsome05dynamictaint}. For example, during fuzzing~\cite{vuzzer, angora}, \dta explores diverse program execution behaviors and checks which input bytes affect the branches (\ie, the taint sinks) of the target program.
In the case of malware analysis~\cite{Yin07panorama}, \dta monitors if program instruction registers (\ie, taint sink) are manipulated by untrusted user input~\cite{Newsome05dynamictaint}. \dta is also applied to identifying user information leakage~\cite{taint_droid}~\cite{sun2016taintart}, where \dta monitors a set of sensitive user data as taint source and a set of sensitive functions (\eg socket write) that leak that data to the outside world as taint sinks.


\dta is typically implemented with taint tags. There are mainly two types of taint tags used in the literature: binary tags and multiple tags. The binary tag approach marks all taint sources with a single binary value: 1 or 0 to represent tainted or untainted respectively. Binary tags are commonly found in simple tasks such as privacy leakage and detecting attacks from user-supplied inputs. However, they fail to monitor more fine-grained information flow used in malware analysis and taint-guided fuzzing because they can only track the existence of taint rather than the ownership of taint source. In contrast, multi-tag \dta, which tracks every taint source independently, tracks taint at a more detailed granularity at the cost of significantly large runtime overhead which grows quadratically with tag size. The large runtime overhead prohibits practical deployment of online \dta tasks to check security properties such as policy enforcement and intrusion detection in Android~\cite{taint_droid}~\cite{sun2016taintart}. Moreover, time-sensitive applications such as fuzzing ideally require taint analysis for a large number of program executions in a short amount of time for defenders to find vulnerabilities before attackers do~\cite{vuzzer}. These limitations are further detailed in~\Cref{app:overhead}.

\medskip
\noindent
\textbf{Fundamental Problems of DTA.} There are three fundamental problems in the design and implementation of taint: under-taint, over-taint, and large runtime overhead. Even with the heavy instrumentations that cause large runtime overheads, manually engineered rules for taint propagation still have poor accuracy at capturing information flow. These limitations of taint severely affect its applicability to real-world programs. A recent work TaintInduce~\cite{taintinduce} has proposed to learn the taint propagation rules instead of manually specifying them. This can increase the accuracy of individual rules, but the error accumulation and large overhead issues still remain, due to propagation-based design. Therefore, we choose to use end-to-end program embeddings, and we conduct influence analysis directly on the neural program to track information flow.

\medskip
\noindent
\textbf{Program Embeddings.}
In general, there are two types of program embeddings, static and dynamic. Static program embeddings first generates a program representation, then use neural networks to encode the representation into embeddings. Example program representations include token sequences~\cite{Gupta2017DeepFixFC, BhatiaS16, Pu2016, devlin2017robustfill}, abstract syntax trees~\cite{Mou2016}, and control and data flow graphs~\cite{xu2017neural,allamanis2017learning}. Static program embeddings have been applied to correcting student code errors, automatic vulnerability detection, and detecting variable misuse. Since such program representations cannot capture program semantics, dynamic program embeddings learn program behavior from input-output pairs~\cite{neural_repair, pmlr-v37-piech15, shen2018neuro} by executing the program. Dynamic program embeddings have been used for fuzzing~\cite{neuzz}, solving symbolic constraints~\cite{shen2018neuro}, program repair~\cite{neural_repair} and generating feedback on student code~\cite{pmlr-v37-piech15}. Since information flow analysis reflects the runtime behavior of programs, we use dynamic program embeddings to learn from program execution traces. Our neural program model approximates the program logic from taint source to taint sink. Then, we analyze the flow of information in the model.


\medskip
\noindent
\textbf{Information Flow in Neural Network.} A popular technique to track information flows in a neural network (\nn) is a saliency map, which measures the sensitivity of the \nn output to changes in the input features ~\cite{Simonyan2013DeepIC}.
For example, in image classification, the saliency map can be viewed as an annotated representation of the input image, where the annotations at every pixel correspond to the gradient of the output \wrt to the corresponding original pixel value ( \ie, how the output category changes as the input image pixels change).
Saliency maps have also been used to construct inputs with minimal perturbations as adversarial examples to an image classifier~\cite{Papernot2016TheLO}. Since the saliency map indicates the most critical input features that affect final neural network output, it guides an attacker's construction of the adversarial example by localizing the changes needed on features to change the classifier output.

As a gradient-based attribution method, a saliency map has been widely used in interpreting neural networks. Compared to other gradient-based methods (\eg integrated gradient~\cite{sundararajan2017axiomatic}), saliency maps focus on the sensitivity of neural output to every feature, \ie, how the \nn output changes with respect to a small change in the input. In contrast, integrated gradient tries to explain the attribution of neural output to every feature, \ie, how each feature of input contributes to the final \nn output. This implies that saliency values for specified input features may differ from their corresponding integrated gradient value. Since integrated gradient value is essentially $gradient * input$ and saliency value is gradient, the disparity between these two values is maximized when the input value is significantly small but the gradient value is large. In our case, since we want to infer which byte in the input affects the taint sink, \ie, induce the greatest sensitivity to the neural network output, we use the saliency map method.
\section{Methodology}

\subsection{Overview}

In this section, we give a motivating example to show the workflow of \tool. As shown on the left side of Fig~\ref{fig:workflow}, we assume the taint source is \texttt{x}, taking 6 bytes from the user input, and the taint sink is variable \texttt{z}. The propagation-based dynamic taint analysis cannot derive accurate information flow in this case. Since variable \texttt{c} at line 4 is computed by \texttt{a} and \texttt{b}, the first two bytes of user input, so taint value for \texttt{c} is \texttt{a} and \texttt{b}. At line 5, \texttt{z} is computed from \texttt{c} and \texttt{b}, thus the taint value for \texttt{z} is composed from \texttt{c} and \texttt{b}.
The analysis is accurate for both line 4 and line 5, but composing the propagation rules together amplifies errors.
The analysis ignores the fact that at line 5 \texttt{z} actually equals \texttt{a*a} and is only affected by the first byte of user input. Composition introduces and amplifies errors and runtime overhead in the dynamic taint analysis. 

On the contrary, \tool uses an end-to-end approach to build neural program embeddings for information flow analysis. Based on some training samples (\ie, user input, \texttt{z}), \tool learns a neural program from dynamic execution results which preserve program context--\texttt{z} is only affected by \texttt{a}. As shown on the right side of Fig~\ref{fig:workflow}, given a user input \texttt{x}, \tool computes the gradient of variable \texttt{z} with respect to \texttt{x} and constructs a saliency map which indicates the sensitivity how each byte of \texttt{x} affects \texttt{z}. From the saliency map, we find that first byte is the most critical byte of input affecting \texttt{z}. Fig~\ref{fig:workflow} presents a high-level overview of our approach.

\textbf{Training.}
We first train a neural program to learn the information flow from taint source to sink. For a given program and a set of inputs, we mark these inputs and use light-weight instrumentation to collect values of sink variables. They represent the dynamic behavior of a program. Next, we train a neural network model (\nn) to learn this dynamic behavior. Our \nn approximates a function that maps sources to sinks. The training process minimizes the errors of learning this function, thereby improving the precision of information flow tracking.

\textbf{Influence Estimation.}
We construct two saliency maps to infer the information flow from taint source to sink. Saliency maps analyze the sensitivity of input features for \tool. The more important a feature is, the more it influences the \nn output. We first define a saliency map to summarize coarse-grained information flow for the program behavior, aggregating gradient information for all inputs and all paths. Then, we define the second saliency map to identify the most important taint sources for specific sinks, utilizing first-order partial derivatives of the \nn output with respect to the input.

Since our end-to-end methodology of collecting program behavior information, training, and influence estimation is lightweight, the runtime overhead is much smaller than a traditional taint analysis tool.
\tool directly performs analysis at the program semantic level by learning from dynamic program behaviors rather than on the instruction semantic level in traditional taint analysis which leads to under-taint and over-taint. Learning the end-to-end model with \tool reduces overall information tracking errors, which mitigates the issues of over-taint and under-taint. Thus, \tool achieves more accurate results than traditional taint analysis tools.

\subsection{Program Embedding} 


\tool learns the information flow by observing a large set of taint source-sink pairs from program execution traces. The model predicts the values of taint sink variables given taint sources as model input. We formally define our neural network model as follows, with detailed architecture shown in Appendix~\ref{app:nn_arch}. Given a set of concrete taint sources $\vx$ and the corresponding taint sinks $\bm{y}$ for a specified program $P$, the neural program predicts the taint sinks as $\bm{\hat{y}}$, with the following equations.

\begin{equation}
\va = \phi(\mW_1^{T}\vx + \vb_1)
\label{eq:1}
\end{equation}
\begin{equation}
{\bm{\hat{y}}} = \sigma(\mW_2^{T} \va + \vb_2)
\end{equation}

We denote $\mW_{k}, \vb_{k}$ as trainable parameters for every layer where $k$ represents the layer index, $\phi$ represents the ReLU function, and $\sigmoid$ represents the sigmoid function. In Equation~\ref{eq:1}, $\va$ represents the output vector of the hidden layer of neural network.
The \nn model learns the function $f$ that takes numerical vector of size $m$ as input and outputs $n$ taint sink variables.
Let $\bm{\theta}$ denote the trainable weight parameters of $f$. Given a set of training samples $(\mX, \mY)$, where $\mX$ is a set of taint sources and $\mY$ represents the correct taint sink values, the training task of the parametric function $f(\vx,
\bm{\theta})$ is to obtain the parameter $\bm{\hat{\theta}}$ that minimizes the multi-variable regression loss, where each variable is a taint sink.

After we train the \nn model, we construct two saliency maps to analyze the flow of information in the neural program. From the neural program model, the first saliency map provides a global view of coarse-grained information flow when all sinks are considered as a whole. The second saliency map can extract the most influential taint sources for any given sink. We now explain the details of the information flow analysis.

\subsection{Coarse-Grained Information Flow}
\label{coarse}

We discuss the method to extract coarse-grained information flow from the \nn model. We define the coarse-grained information flow as the influence of each source on \emph{all} sinks. Since some dynamic taint analysis applications have a set of taint sink variable, e.g., taint-guided fuzzing, it is important to consider coarse-grained information flow to all the sinks. The aggregated information flow to a set of taint sink variables can highlight which part of the taint source has the most significant effect on them.

To extract coarse-grained information flow, we first compute the  partial derivatives of the taint sink with respect to all sources. Let $f_i(\bm{\theta},\bm{x})$ denote the output value for the $i$-th taint sink variable during the execution of the targeted program with taint source $\bm{x}$. We compute the derivative with respect to a given taint source $\bm{x}$, defined below, where $x_j$ denotes the j-th byte in the taint source.
\begin{equation}
\label{eqn:j_m}
\nabla_{\bm{x}}f(\bm{\theta}, \bm{x})= \frac{\partial f(\bm{\theta}, \bm{x})} {\partial \vx} = \bigg[ \frac{\partial f_i(\bm{\theta}, \bm{x})} {\partial x_j} \bigg]_{i\in1...n, j\in1...m}
\end{equation} 

The partial derivatives constitute a Jacobian matrix of the neural network function. Each element of the matrix represents the gradient of output neuron $f_i(\bm{\theta}, \bm{x})$ with respect to taint source byte $x_j$. Note that the gradient we compute has two main differences from the gradient used in a neural network trained by backpropagation. First, the target function is different. The gradient used for backpropagation is computed on a loss function which includes information about the state of the model parameters and the expected outputs. In contrast, our method computes the derivative on the output of the neural network, which includes only information about the model parameters. Since we aim to interpret how neural networks make the decision after convergence, our gradient computation does not need to consider the corresponding ground truth information. Second, we compute the gradient with respect to the input, rather than trainable parameters of neural network model. By computing the gradient directly with respect to the input, we obtain the sensitivity of \nn output to all the bytes in the input.


Then, we construct a saliency map to provide the global view for coarse-grained information flow, based on partial derivatives of the neural network model. The saliency map $S(\bm{x})$ is defined as follows.
\begin{equation}
	S(\bm{x})[j] = \sum\limits_{i}\bigg| \frac{\partial f_i(\bm{\theta}, \bm{x})} {\partial x_j}\bigg|
\label{n_sink}
\end{equation}
$S(\bm{x})[j]$ is the sum of all the sink sensitivity to the j-th byte, representing the effect of the j-th byte to the overall program behavior from the current execution. Summarizing the sensitivity to all sinks includes information about all paths to these sinks. In addition, the neural program includes information about all the input data. Therefore, we can analyze the coarse-grained information flow using this saliency map.

\subsection{Fine-Grained Information Flow}
\label{sel_sink}
We define fine-grained information flow as the influence of each source to a \emph{single} sink.
Dynamic taint analysis applications that are interested in fine-grained information flow often set taint sink at a certain variable such as function pointer, jump target address and instruction pointer register. We refer to these applications as fine-grained information flow analysis.

To reason about how information arrives at a given sink, we follow similar steps from the coarse-grained information flow analysis as mentioned in~\Cref{coarse}. First, we compute the Jacobian matrix to obtain the gradient information using equation~\ref{eqn:j_m}. After obtaining the gradient value for every byte in the taint source, we can construct a saliency map to infer the fine-grained information flow from taint source to a particular taint sink. Since we are only interested in the sensitivity of the taint sink to every byte in the taint source, we take the absolute value of the gradient to construct the saliency map $S(\bm{x})$ defined as follows. 
\begin{equation}
	S(\bm{x})[j] = \bigg| \frac{\partial f_i(\bm{\theta}, \bm{x})} {\partial x_j}\bigg|
\label{1_sink}
\end{equation}
The bytes that causes the maximum fluctuations of \nn output are considered taint source bytes that influence the sink. 
The set of source bytes that determines taint sink variables can be inferred by finding the top-k bytes with maximum values, defined  below.   

\begin{equation}
    H_i(k): arg({top\_k}(\bigg| \frac{\partial f_i(\bm{\theta}, \bm{x})} {\partial \bm{x}}\bigg|))
\label{top_k}
\end{equation}
Let $H_i(k)$ denote the set of indices of K source bytes, $top\_k$ denotes the function to select $k$ largest elements from a vector and $arg$ denote the function to return indices of selected elements. Since our neural network learns a summary of dynamic program behavior from all training samples, the influential source bytes inferred from the neural network model contains knowledge from a large number of concrete runs of the program. On the contrary, traditional dynamic taint analysis tools have information from only the specific path taken from one execution.

\subsection{Data Collection}

In this section, we describe the general method to collect a set of training samples for our neural program training. To learn the information flow from taint source to taint sink, it is crucial to obtain a large, diverse, and representative dataset. However, unlike traditional machine learning tasks (\eg image classification, natural language processing, speech recognition), there is no standard dataset for various taint sources and sinks. There are many options to collect training datasets. A natural solution to collect such a dataset would be to randomly sample taint source-sink execution pairs for a specified programs. As an example to generate a training dataset, we can start with a common taint source, randomly flip the bytes in the taint source, and record the corresponding taint sink values. Alternatively, we can use a simple fuzzer to generate a set of taint source which trigger diverse program states and record the taint sinks values. The training data coverage affects the amount of information \tool can track. We can further improve the information flow coverage using more sophisticated techniques like coverage-guided fuzzing, symbolic execution, etc. However, in this paper we demonstrate that even with training data generated by a simple fuzzer, \tool can easily outperform existing DTA tools.

Note that taint sources are normally user input, files or user privacy strings that can be represented as byte sequences. So we can easily convert the byte sequences to bounded numerical vectors ranged in $[0, 255]$. However, the taint sink can be arbitrary variables in the program with unbounded values such as instruction pointer register, a complex socket structure, or a user defined variable in the program. These arbitrary variables are hard to model as unified representations for \nn output and make it difficult for the \nn to converge. To tackle this problem, we normalize these unbounded variables to bounded data for different applications. For example, in taint-guided fuzzing, we set taint sinks at a set of variables used in branch conditions and normalize the sink variables with binary data (\ie, 1 represents the branch is taken, 0 represents the branch is not taken). The binary representation of the \nn output can ensure the fast convergence of the model.

\section{Evaluation}
In this section, we evaluate the effectiveness and efficiency of \tool against three state-of-the-art dynamic taint analysis tools (Libdft~\cite{libdft}, Triton~\cite{triton}, and DFSan~\cite{dfsan}). We answer the following research questions about \tool in the evaluation.
\begin{enumerate}
    \item \textbf{Hot Byte Accuracy:} Is \tool more accurate at finding hot bytes (\ie, the most influential bytes that determine different program behaviors) in input for 6 real-world programs?
    \item \textbf{Runtime Overhead:} What is the runtime overhead of \tool compared to state-of-the-art dynamic taint analysis tools  (both with and without GPU)?
    \item \textbf{Exploit Analysis:} Can \tool detect vulnerabilities in real-world programs?
    \item \textbf{Application on Taint-Guided Fuzzing:} Since fuzzing is one of the most important security applications of taint, does \tool help taint-guided fuzzing achieve better edge coverage compared to other taint analysis tools?
    \item \textbf{Model Choice:} How does \tool perform with different machine learning models other than neural networks?
    \item \textbf{Information Loss:} What kinds of flows are missed by \tool? How does the training data quality affect such information loss and how to mitigate this loss?
\end{enumerate}

To answer these questions, we will first describe our experiment setup and how we learn the neural program embeddings for the real-world programs.


\subsection{Experiment Setup}
\label{sec:exp_setup}
\medskip
\noindent
\textbf{Environment Setup.} All our measurements are performed on a ubuntu 16.04 system with an Intel Xeon E5-2623 v4@2.60GHz CPU , an Nvidia GTX 1080 Ti GPU and 256 GB RAM. We implement the \tool in Keras-2.1.4~\cite{keras} with Tensorflow-1.8.0~\cite{tensorflow} as the backend. 

Next, we give a brief introduction of the three front-end tools in our evaluation, how we set up the tools, and our implementation of \tool.

\subsubsection{Libdft}
\noindent
Libdft is a widely-used dynamic taint analysis engine based on Intel PIN framework~\cite{pin}. It predefines taint propagation rules for every type of instruction via external functions using the PIN analysis API. Then, it dynamically instruments the binary code using the PIN instrumentation API at runtime. For every executed instruction, Libdft calls a corresponding external function to track the taint flow. If a particular type of instruction (\eg pop, ret) lacks any taint information, no external function will be invoked. Nevertheless, for real-world applications, most instructions contain taint information, and hence this design incurs a large runtime overhead due to the sheer number of external function calls. Another notable drawback of Libdft is that the current implementation only supports the x86 architecture. 

\medskip
\noindent
\textbf{Setup.}
We set up a modified version libdft with support of multiple taint tags~\cite{libdft_multi}. The dependency PIN version is 2.13.
\subsubsection{Triton}

\noindent
Triton is a platform that supports concolic execution, dynamic taint analysis, and abstract syntax tree representation. Similar to Libdft's approach for dynamic analysis, it uses Intel PIN to monitor taint flow corresponding to a set of predefined taint propagation rules. Currently, its taint analysis engine only supports the x86 architecture. Triton provides users Python bindings to the underlying PIN API so that they can write scripts to perform customized analysis tasks. However, these bindings are limited and fail to capture the full functionality of PIN. Moreover, the limited Python bindings cause imprecise dynamic taint analysis results in addition to the runtime overhead from heavy instrumentation and monitoring taint propagation in PIN.  

\medskip
\noindent
\textbf{Setup.}
We set up a develop fork Triton to support the multiple taint tags feature~\cite{triton_multile}. The dependency PIN version is 2.13. We write the analysis script with Triton Python binding to set corresponding taint sources and taint sinks according to different programs.
\subsubsection{DFSan}

\noindent
DFSan (DataFlowSanitizer) is a data flow analysis framework provided by Clang~\cite{clang}. It consists of a compile-time instrumentation module and a runtime dynamic library to track taint flow for the x86-64 architecture only. Users only need to define taint source and taint sink with the public DFSan API. DFSan relies on predefined taint propagation rules for LLVM IR instructions rather than architecture-specific assembly instructions. This enables DFSan to insert taint tracking functions at compile time into a program. Thus, it has a smaller runtime overhead compared to other PIN-based tools that use dynamic instrumentation. DFSan, however, fails to run on programs which depend on external shared libraries. Since the dynamically shared libraries cannot be instrumented when compiling a given program, DFSan cannot insert these taint tracking functions and fails to work on programs depending on dynamic shared libraries. This along with the exhaustion and resolution of taint tags (\Cref{sec:eval_overhead}) limits the applicability of DFSan to real-world applications. 

\begin{table}
\footnotesize
\centering
\begin{tabular}{cccc}
\toprule
{\bf DTA Engine} & {\bf Propagation Level} & {\bf Dependency} &{\bf Tag Type}\\
\midrule
Libdft & assembly instruction & Pin 2.13 & multi-tag \\
Triton & assembly instruction & Pin 2.13 & multi-tag \\
DFSan & LLVM instruction & LLVM-7.0.0  & multi-tag \\
\bottomrule
\end{tabular}
\caption{Dynamic Taint Analysis Engines.}
\label{stdied_dta}
\end{table}

\medskip
\noindent
\textbf{Setup.}
We set up DFSan from the Clang runtime library. The underlying LLVM version is 7.0.0. We use DFSan's API to set taint source and sink for different programs.
\noindent
\subsubsection{\tool}

\medskip
\noindent
\textbf{Model Architectures.}
For each program, we train a neural program model which learns the program logic from taint sources to taint sinks. The \nn model consists of $3$ fully-connected layers. The hidden layer uses ReLU as activation functions with $4096$ hidden units. The output layer uses Sigmoid as the activation function to predict the sink variables. Since each program has different taint sources and taint sinks, the corresponding neural program model has different number of input/output neurons. We describe taint sources and taint sinks for all programs in Table~\ref{tab:taint_num}. We use the first 6 programs to evaluate the hot byte accuracy and the taint-guided fuzzing experiments, so we set multiple taint sinks as branch variables (\ie, the variables used in conditional predicates). The later 5 programs are evaluated in the exploit analysis experiment, so they only have a single taint sink at a specified variable. All 11 programs set each byte of the program input as a taint source. Thus the total number of taint sources are the total number of input bytes.

\medskip
\noindent
\textbf{Training Data Collection.}
To collect the training data, we first run the AFL fuzzer with an initial seed to collect its mutation corpus. Next, we use a simple LLVM pass to add light-weight instrumentation for recording the two operands of CMP instructions during runtime. These operands of CMP instructions are branch variables (\ie, our taint sinks) evaluated in conditional predicates. 
We run the instrumented program with the generated input and record the taint sink values. We collect around 2K input-output pairs (each input has multiple source bytes and reaches multiple sinks) on each program for training. For hot byte evaluation, we normalize the taint sink variables into binary, \ie, 1 if a sink value satisfies the conditional predicate and 0 if not. We check the value of predicates by computing the difference of two operands of CMP instructions. For exploit analysis, we use the similar LLVM pass to obtain the specified taint sink values, then perform standard min-max normalization to the sink values (\ie, $y_{norm} = (y - y_{min})/(y_{max} - y_{min})$). Note that the data collection and normalization can be easily done by a simple python script, no manual labeling effort is required.

\medskip
\noindent
\textbf{Training Procedure.}
We adopt random weight initialization and cross-entropy loss function or mean-square-error loss depending on specific model output data type. The NN model is trained with an Adam optimizer for $100$ epochs with an initial learning rate $0.01$ and decay rate $0.7$ per $10$ epochs. We choose a mini-batch size $16$. For exploit analysis evaluation, we use mean-square-error as loss function and metric to evaluate our model performance. Since our \nn model is simple, the training process is very efficient and takes on average 73s across all tested programs. In~\Cref{hot_byte_eval}, we test the accuracy and false positive rate of our neural program models at identifying hot bytes. In addition, we evaluate the information loss in~\Cref{flowloss}.

\begin{table}
\footnotesize
\centering
\begin{tabular}{lcll}
\toprule
{\bf Program} & {\bf Taint Sources \#} & {\bf Taint Sinks \#} & {\bf Taint Sinks Types} \\
\midrule
readelf & 7467 &  2122 & branch variables\\
harfbuzz & 5049 & 2805 & branch variables\\
mupdf & 4861 & 1377 & branch variables\\
libxml & 8040 & 1929 & branch variables\\
libjpeg & 5873 & 997 & branch variables\\
zlib & 8306 & 571 & branch variables\\
\midrule
sort & 500 & 1 & length variable\\
openjpeg2 & 298 & 1 & denominator\\
libsndfile & 446 & 1 & length variable\\
nm & 847 & 1 & counter variable\\
strip & 1123 & 1 & length variable\\
\bottomrule
\end{tabular}
\caption{We set program input bytes as taint sources, and we use different types of taint sinks. For the first 6 programs, we select sink variables at program branches to evaluate \tool's accuracy, overhead, and application on taint-guided fuzzing. For the rest 5 programs, we set taint sinks according to the vulnerability information for exploit analysis.}
\label{tab:taint_num}
\end{table}

\subsection{Is \tool more accurate at finding hot bytes (\ie, the the most influential bytes) in input for 6 real-world programs?}
\label{hot_byte_eval}

\begin{table*}[t]
\scriptsize

\begin{center}
\begin{tabular}{lc|cccccccc|cccc}
\toprule

\multirow{2}{*}{\bf Programs}&{\bf File}& \multicolumn{3}{c}{\bf Hot Byte Accuracy}& & &  \multicolumn{2}{c}{\bf Hot Byte FPR}& &  \multicolumn{4}{c}{\bf Runtime} \\
& \bf Format & \bf \tool & \bf Libdft & \bf DFSan & \bf Triton & \bf \tool & \bf Libdft & \bf DFSan & \bf Triton & \bf{\tool} & \bf Libdft & \bf DFSan & \bf Triton\\
&  &  &  &  &  &  &  &  &  & \bf (GPU/CPU) &  &  & \\
\midrule
readelf-2.30 & ELF & 81\% & 74\%  &49\%  & 44\%$^\dagger$ & 1.6\% & 3.0\% & 3.5\% &  3.7\%$^\dagger$ & 6m/25m & 161m & 117m & $>$24h\\
harfbuzz-1.7.6 & TTF & 86\% & 86\% & n/a  & 13\%$^\dagger$ & 0.8\%& 0.8\% & n/a & 5.2\%$^\dagger$ & 5m/18m & 204m & n/a & $>$24h \\
mupdf-1.12.0 & PDF & 80\% & 48\% & 57\% & 33\%$^\dagger$ & 2.0\% & 2.9\% & 2.6\% & 3.9\%$^\dagger$ & 3m/9m & 224m & 532m & $>$24h \\
libxml2-2.9.7 & XML & 65\% & 56\% & n/a & 14.2\%$^\dagger$ & 2.3\% &  3.0\% & n/a & 5.8\%$^\dagger$ &  6m/29m & 197m & n/a & $>$24h \\
libjpeg-9c  & JPEG & 29\% & n/a & n/a & n/a &  3.9\% & n/a & n/a & n/a & 3m/5m & n/a & n/a & n/a\\
zlib-1.2.11 & ZIP &  66\% & 26\% & n/a & 3\%$^\dagger$ & 1.8\% &  3.9\% & n/a & 5.1\%$^\dagger$ &3m/4m & 20m & n/a & $>$24h \\
\bottomrule
\end{tabular}
\end{center} 
\hspace{0.1cm}{\scriptsize \dag indicates cases where Triton analyzed partial dataset within 24 hours}
\caption{For each program, we measure the accuracy and false positive rate of identifying hot bytes, as well as runtime overhead for different DTA tools and \tool. \tool achieves the highest accuracy and lowest false positive rate. On average, \tool increases the accuracy by 10\% and reduces the false positive rate by 0.44\% compared to the second-best DTA tool Libdft. \tool is $40\times$ faster on GPU and $10\times$ faster on CPU than the second-best DTA tool Libdft.}
\label{hot-accur}
\end{table*}
It is extremely hard to evaluate the accuracy of dynamic taint analysis. We would like to find ground truth for taint that is relevant to its applications.
DTA is often used to search for important bytes in the inputs to trigger specified program behaviors. For example, vulnerability analysis needs to find which part of untrusted input triggers malicious behaviors. Taint-guided fuzzing aims to find importance bytes which explore new program behaviors and yield new code coverage. DTA finds these important bytes by setting and propagating taint labels from taint sources (program input) to taint sinks (various variables that determines program behaviors). 
Therefore, we propose to use these \emph{hot bytes} to evaluate the accuracy of DTA tools and \tool. 

\medskip
\noindent
\textbf{Hot Bytes Definition.}
We define importance bytes that can maximally influence the variables in program branches as \emph{hot bytes}.

Then next question is how to obtain the ground truth hot bytes as baseline to compute the hot byte accuracy. We observe that a large number of real-world programs are parsers which take in a specified file type and check its format. Meanwhile, most of the program behaviors of these parser programs are determined by bytes at the specified locations of input (\ie, the fixed locations where file format headers locate), rather than the file content. Hence, for a parser program, we can approximate that the hot bytes are mostly located at the structured format sections. Further, by analyzing the fixed structured locations of a file, we can obtain the estimated ground truth of hot bytes for a particular parsing program. These ground truth hot bytes can be used as a metric to evaluate the effectiveness and efficacy of dynamic taint analysis tools on a particular parsing program.

\medskip
\noindent
\textbf{Ground Truth.}
Note that all evaluated real-world programs in this sections are file parsing programs. Since these $6$ parsing programs have particular structures in their file formats, we can obtain estimated ground truth of hot bytes by analyzing these file formats. A saliency map of ELF file format is shown in Appendix~\ref{app:saliency}.
Fig~\ref{heat_map} shows the ground truth of all 6 file formats. Similar to ELF, other formats also include the header and trailers. But some files may have unique format features. For example, ZIP file has an additional local file header after the ZIP file header at the beginning. TTF file contains unique character tables near the trailers. 

\medskip
\noindent
\textbf{Extract Hot Byte.}
We perform the following step to extract the hot bytes from our neural program models. 1) 
For each program, we feed the seed input to \nn model. 2) We compute the gradient of taint sinks with respect to the taint sources (seed input) and construct the saliency map using Equation~\ref{n_sink}. The saliency value indicates the extent to which the byte in taint sources affect the all the taint sink variables. 3) We select the top 5\% bytes with highest values from the input saliency map as possible hot bytes using Equation~\ref{top_k}. The reason we set this threshold is that in practice, only a small number of hot bytes in the input determine program behaviors. After analyzing all $6$ file formats, we find that the total number of ground truth hot bytes range from 250 to 500 which takes around 5\% of total input bytes.

\medskip
\noindent
\textbf{Compute Hot Byte Accuracy.}
We compute the hot byte accuracy by checking if the hot bytes analyzed from taint tools are consist with ground truth hot bytes. To be specific, if a hot byte identified by \tool locate in the estimated ground truth range, we consider it as a true hot byte identification (\ie, true positive); otherwise, we consider it as false identification (\ie, false positive). 
For the 3 other state-of-the-art dynamic taint analysis tools, we also evaluate their abilities to find hot bytes to compare against \tool. Since dynamic taint analysis operates on a single execution trace, we run the dynamic taint analysis tools on every input in the training data and collect the tainted bytes for every execution; we then aggregate these tainted bytes by counting the total number of times a specified byte is tainted. In this way, we construct a similar saliency map as \tool for these 3 tools. We then select the same threshold 5\% of top tainted bytes as possible hot bytes and calculate the hot byte accuracy. 
Lastly, we measure the total runtime to obtain the final hot byte accuracy for the 4 tools. For the 3 dynamic taint analysis tools, we record the total runtime for tracking all the input samples in the dataset. For \tool, we record the total runtime cost for \nn training and gradient computation.
Based on the best-effort ground truth for the 6 file formats, we compute accuracy and false positive rate of identifying hot bytes in a standard way.

\begin{figure*}[t!]
\begin{center}

\subfloat[ELF]
{\includegraphics[width=0.33\textwidth]{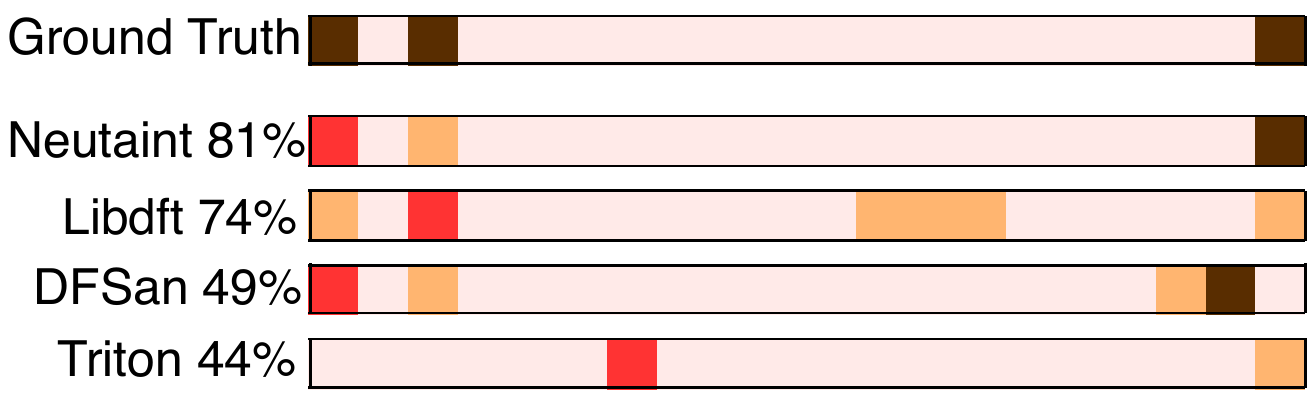}}
\hfill
\subfloat[XML]
{\includegraphics[width=0.33\textwidth]{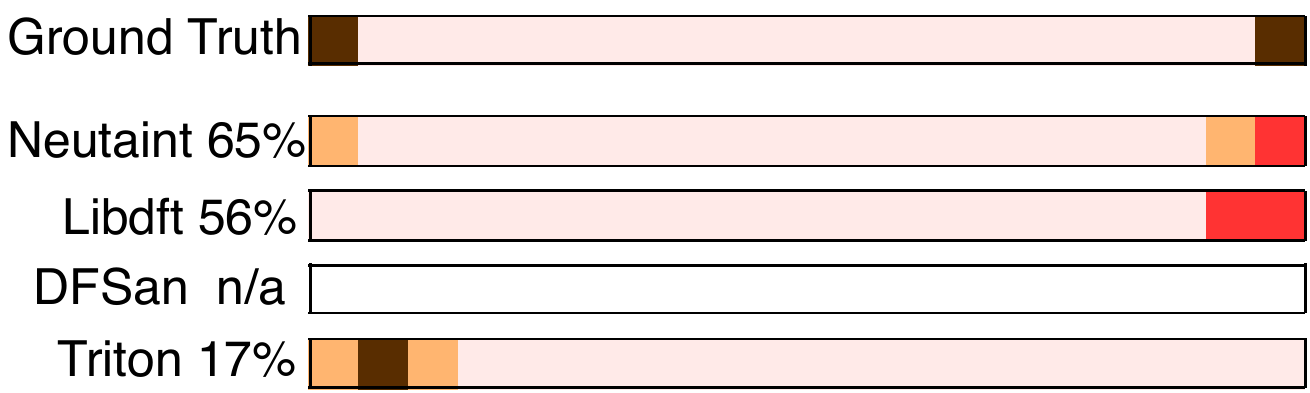}}
\subfloat[TTF]
{\includegraphics[width=0.33\textwidth]{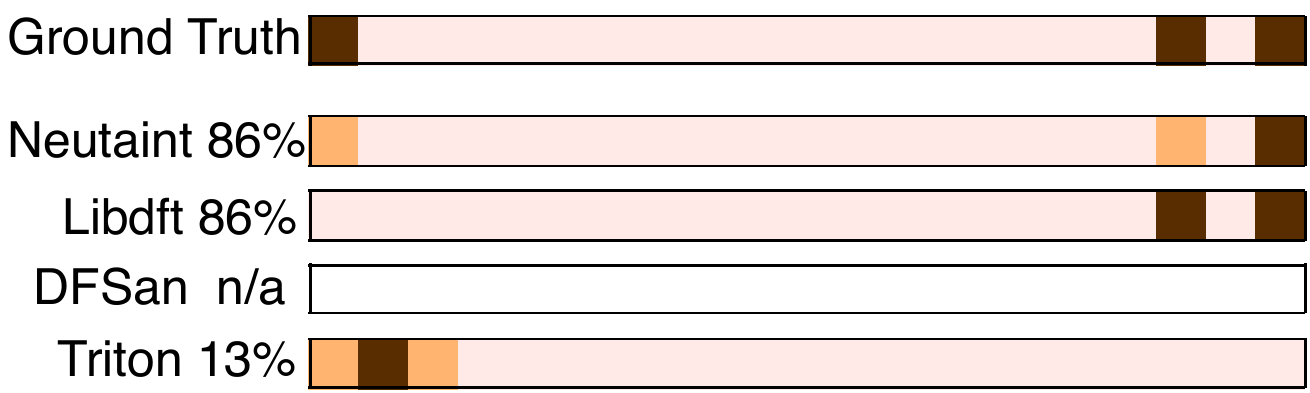}}

\subfloat[JPG]
{\includegraphics[width=0.33\textwidth]{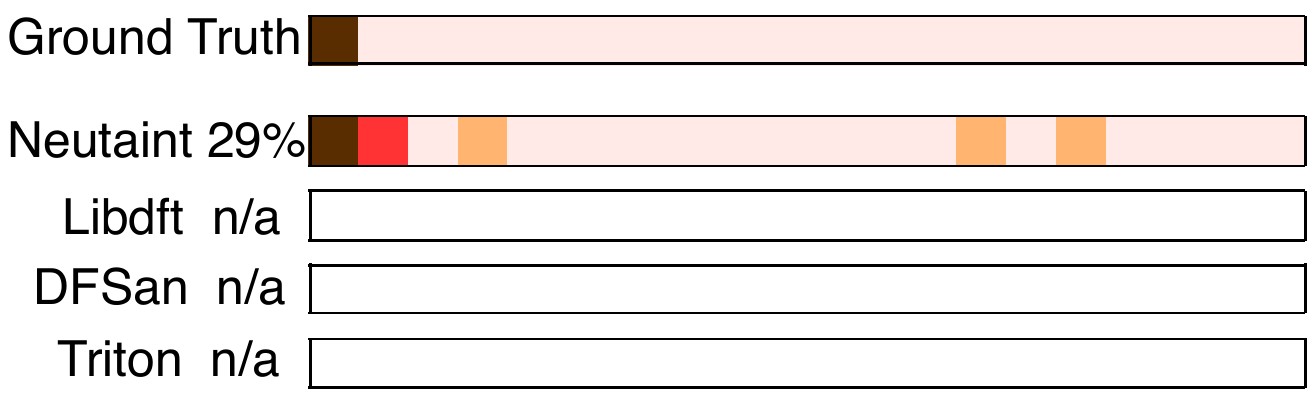}}
\hfill
\subfloat[ZIP]
{\includegraphics[width=0.33\textwidth]{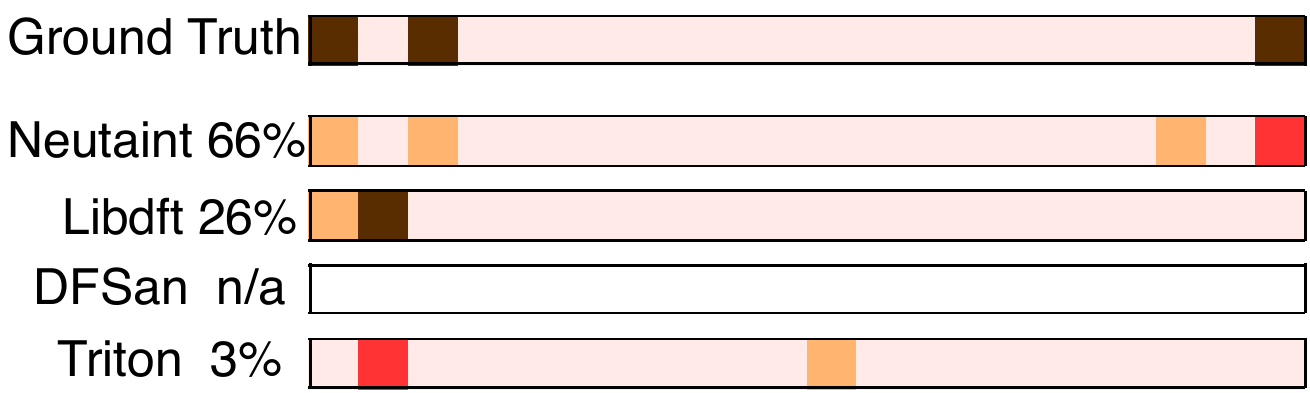}}
\subfloat[PDF]
{\includegraphics[width=0.33\textwidth]{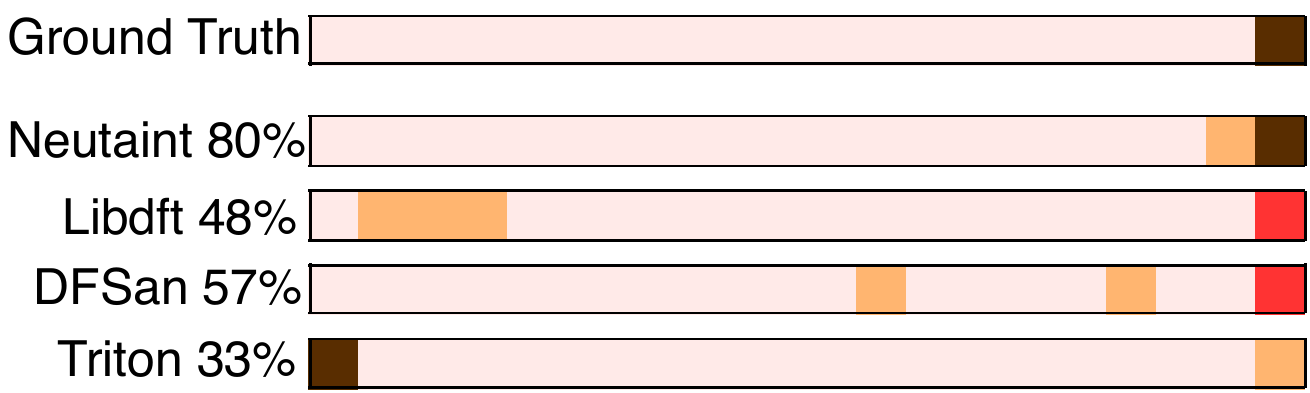}}
\end{center}

\vspace{-0.4cm}
\hspace{1.2cm}\subfloat
{\includegraphics[width=0.25\textwidth]{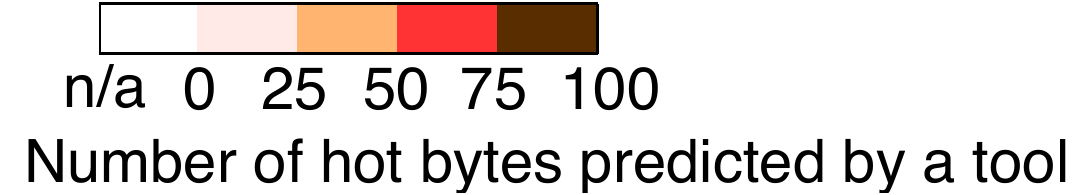}}
\caption{The six heatmaps show how each tool identifies hot bytes for a given file format. The x-axis is broken into byte intervals, and an interval's darkness is proportional to how many hot bytes a tool predicts. Since the first row represents the ground truth, the correctness is defined by how each subsequent row aligns with the ground truth row.}
\label{heat_map}
\end{figure*}

\medskip
\noindent
\textbf{Results.} 
The results for hot byte accuracy is shown in Table~\ref{hot-accur}. For programs with simple and straightforward parsing logic such as \texttt{readelf} and \texttt{harfbuzz}, we observe that all four tools find hot bytes with high accuracy. For programs with complex parsing and transformation logic such as \texttt{libjpeg} and \texttt{zlib}, the accuracy drops for all tools due to large taint flows inside the decompression algorithm. Nonetheless, \tool achieves the highest hot byte accuracy on $5$ programs. Even for the remaining program \texttt{libjpeg} which causes significant runtime overhead due to large taint propagation flows, \tool is the only one to be able to finish the analysis within a reasonable time (minutes as opposed to hours). Fig~\ref{heat_map} shows the detailed visualisation of hot byte accuracy using a heat map. 

Traditional taint propagation can result in significant slowdowns for time-sensitive operations in large real-world applications. \texttt{libjpeg}'s decompression algorithm requires massive memory read and write operations which carry taint information. Indeed, all three traditional dynamic taint analysis tools fail to finish the analysis of \texttt{libjpeg} on the dataset within 24 hours. Moreover, the second-best tool Libdft spends more than 3 minutes on \texttt{libjpeg} to finish a single execution (more than 2 days for all $993$ inputs in the dataset) to track all the taint flows inside the decompression algorithms while a vanilla execution of \texttt{libjpeg} without dynamic taint analysis takes $0.015$s. The runtime overhead for intense taint tracking could be more than $10^4$X times while \tool avoids such heavy instrumentation overhead through a lightweight neural network model that identifies the hot bytes. 

\medskip
\noindent
\textbf{Analysis: Hot Byte Accuracy.} 
The average hot byte accuracy for all 4 tools across all evaluated programs are $68\%, 58\%, 53\%, 27\%$. \tool achieves $10\%$ more accuracy improvement than second-best dynamic taint tool Libdft. As for hot byte accuracy of every program, \tool achieves $7\%, 0\%, 7\%, 23\%, 40\%$ more improvement respectively on program \texttt{readelf}, \texttt{harfbuzz}, \texttt{mupdf}, \texttt{libxml}, \texttt{zlib}. The reason for \tool's higher accuracy is that \tool is an analysis based on program semantics by learning dynamic program logic rather than an analysis based on instruction semantics performed by traditional dynamic taint analysis tools. \tool could flexibly adapt to diverse execution context which cannot be accurately modeled by dynamic taint analysis tools through fixed, predefined taint propagation rules.

\medskip
\noindent
\textbf{Analysis: Hot Byte False Positive Rate.} 
For all programs, \tool achieves the lowest false positive rate at identifying hot bytes (Table~\ref{hot-accur}). On average,
\tool has 2.07\% false positive rate, that is less than half of that for Triton. Compared to the second-best DTA tool Libdft, the hot byte false positive rate of \tool is 0.44\% lower. The results show that learning end-to-end program embeddings can effectively reduce the overtaint issue.

\smallskip
\begin{longfbox}
\textbf{Result 1:} \tool achieves the highest hot byte accuracy and lowest false positive rate in six popular file formats compared to state-of-the-art dynamic taint analysis tools. On average, \tool increases the accuracy by 10\% and reduces the false positive rate by 0.44\% compared to the second-best DTA tool Libdft.
\end{longfbox}

\subsection{What is the runtime overhead of \tool compared to state-of-the-art dynamic taint analysis tools, with and without GPU?}
\label{sec:eval_overhead}
We measure the runtime overhead for \tool and all three dynamic taint analysis tools. We measure the total time needed to process all the inputs in our dataset, as the runtime in Table~\ref{hot-accur}. For \tool, the total runtime of \tool is composed of three parts, collecting program behavior data, training \nn and computing the saliency map of \nn.

Table~\ref{hot-accur} summarizes the result.
Overall, \tool has the least runtime overhead for all six programs compared to other tools, since the time cost for training and computing the saliency map is negligible. To collect the training dataset, we obtain values of the sink variables in the binary through light-weight instrumentation which introduces negligible overhead. In addition, computing the saliency map is computationally efficient. Therefore, \tool enjoys the fastest runtime among all four tools evaluated. In particular, on program \texttt{mupdf}, \tool can save up to $74\times$ runtime overhead on GPU and $24\times$ runtime overhead on CPU than the second-best DTA tool Libdft. The average runtime overhead for the four tools are $4$ mins(GPU)/$15$min (CPU), $161$ mins, $325$ mins and $>24$ hours. Compared to the second fastest tool Libdft, \tool on average achieves $40\times$ and $10\times$ smaller runtime overhead on GPU and CPU, respectively.

Among the other three tools, Triton achieves the worst result on hot byte accuracy and runtime overhead, due to PIN dynamic instrumentation and inefficient analysis routine. So we only evaluate the partial inputs from dataset on Triton within 24 hours to compute the hot byte accuracy. As for runtime of Triton, we use $>$ 24 hours to indicate the significantly large runtime overhead.
DFSan achieves the second fastest analysis on program \texttt{readelf}. Its runtime overhead is smaller than Libdft because of its efficient instrumentation during compile time rather than runtime. However, DFSan fails to run on four programs because it cannot instrument external dynamically linked libraries at compile time, and it incurs large overhead for recursive tag resolution.
Libdft runs faster than DFSan and Triton for all programs except \texttt{readelf}. It is also more accurate at identifying hot bytes than DFSan and Triton, in all programs except \texttt{mupdf}. However, Libdft is still 43 times slower than \tool, due to execution of heavily instrumented binary and accumulated overhead through propagation.

\begin{longfbox}
\textbf{Result 2:} The runtime overhead of \tool is $40\times$ faster on GPU and $10\times$ faster on CPU than the previously fastest dynamic taint analysis tool Libdft.
\end{longfbox}

\medskip
\noindent
\textbf{Ablation Studies.} We break down the total runtime overhead of \tool for processing 2,000 inputs into three parts, data collection, training and saliency map on both GPU and CPU settings. The results are shown in Table~\ref{tab:runtime_b}. The average runtime of \tool is $244$s across 6 programs, $4\times$ faster than on CPU ($898$s). Even on a machine with only CPU computation, the runtime overhead of \tool is still $10\times$ lower than traditional DTA tools. Since our \nn model has a small number of hyperparameters, the model can be efficiently trained with and without GPU. With GPU, the training time takes around $60\%$. The construction of saliency map and data collections takes around $30\%$ and $10\%$, respectively. With only CPU, the runtime splits into training, saliency map computation, and data collection as $88\%$, $9\%$ and $3\%$. Using only CPU causes on average $4.5\times$ and $0.16\times$ slowdown in training and saliency map construction than with GPU, respectively. In general, the training time takes up the majority of total runtime. Whereas, data collection takes up the least part of runtime because the light-weight instrumentation of program only records the taint sink values during execution and introduces minimal overhead than vanilla execution.

\begin{table}
\scriptsize
\centering
\begin{tabular}{lcllllll}
\toprule
\multicolumn{8}{c}{\textbf{All 2,000 Inputs}} \\
\hline
\multirow{2}{*}{\bf Program} & \bf Data & \multicolumn{2}{c}{\bf Training} & \multicolumn{2}{c}{\bf Saliency Map} & \multicolumn{2}{c}{\bf Total}\\
& \bf Collection & GPU & CPU& GPU & CPU& GPU & CPU\\
\midrule
readelf & 18s & 214s &1352s & 98s&119s & 330s&1471s\\
harfbuzz & 40s & 115s&951s & 119s&130s & 274s&1080s \\
mupdf & 30s & 116s&472s & 64s&82s & 210s&554s \\
libxml & 44s & 216s&1652s & 93s&109s & 353s&1761s \\
libjpeg & 10s & 112s&251s & 36s&37s & 158s&288s\\
zlib & 5s & 110s&202s & 26s&30s & 141s&232s\\
\bottomrule
\end{tabular}
\caption{Runtime breakdown for \tool. The table shows the total runtime of processing 2,000 inputs for data collection, training, and saliency map computation.}
\label{tab:runtime_b}
\end{table}
\begin{longfbox}
\textbf{Result 3:} The total runtime for \tool to process all 2,000 inputs is $244$s on average for each of the six programs. Training time takes up the most part of total runtime overhead, around $60$\% with GPU and 88\% without GPU. 
\end{longfbox}

\subsection{Can \tool detect vulnerabilities in real-world programs?}
\begin{table}
\small
\centering
\begin{tabular}{lcl}
\toprule
{\bf Program} & {\bf Vulnerability Type} & {\bf CVE ID }\\
\midrule
sort & buffer overflow & CVE-2013-0221 \\
openjpeg2 & integer division-by-zero & CVE-2016-9112 \\
libsndfile & out-of-bound read & CVE-2017-14245 \\
nm & heap overflow & CVE-2018-19931 \\
strip & integer overflow & CVE-2018-19932 \\
\bottomrule
\end{tabular}
\caption{\tool can successfully identify the information flow from source to sink in the following exploits.}
\label{tab:cve}
\end{table}

We evaluate the effectiveness of \tool in the analysis of software attacks. We choose 5 known real-world vulnerabilities as listed in Table~\ref{tab:cve}. These vulnerabilities are all from open-sourced programs. Then we perform light-weight instrumentation on the programs and record the values for (taint source, taint sink) pairs during runtime. The 5 vulnerabilities covers various exploit types such as buffer overflow, heap overflow, integer overflow, integer division-by-zero and out-of-bound read. For each vulnerabilities, we set the taint sink at the variables which causes the vulnerability (\eg length variable used by read/write function, variable used as denominator). We also set program input as taint source for every vulnerability. To collect the training samples, we randomly flip bytes at a specified program input (the exploit) and execute the vulnerable programs with the generated input to record the corresponding taint sink values. For the 5 vulnerabilities, we generate 2K training samples for each of them. We train the neural program model which learns the mapping from taint source to taint sink. Then we feed the exploit as taint source to the neural program and construct the saliency map using Equation~\ref{1_sink} based on the gradient of taint sink with respect to taint source. According to the saliency, we can infer which part of taint source determines the taint sink variables. The key result is that \tool successfully locates the hot bytes which control the taint sink variables, and thus can reason about the influence from taint source to taint sink.

\begin{longfbox}
\textbf{Result 4:} \tool can successfully find the information flow from source to sink in known CVEs.
\end{longfbox}

\subsection{Since taint-guided fuzzing is one of the most important security applications of taint, does \tool help taint-guided fuzzing achieve better edge coverage compared to other taint analysis tools?}
\label{subsec:fuzzing}
\begin{table}[t]

\footnotesize
\begin{center}
\begin{tabular}{lccccc}
\toprule
\multirow{2}{*}{\bf Programs}&{\bf File}&\multicolumn{4}{c}{\bf Edge coverage} \\
& {\bf Format} & \bf \tool & \bf Libdft & \bf DFSan & \bf Triton\\
\midrule
readelf-2.30 & ELF & 5540 & 4164 & 2489 & 440$^\dagger$ \\
harfbuzz-1.7.6 & TTF & 5395 & 3796 & n/a & 11$^\dagger$ \\
mupdf-1.12.0 & PDF & 399 & 248 & 192 & 48$^\dagger$ \\
libxml2-2.9.7 & XML & 918 & 428 & n/a & 236$^\dagger$ \\
libjpeg-9c  & JPEG & 649 & n/a & n/a & n/a \\
zlib-1.2.11 & ZIP & 200 & 131 & n/a & 54$^\dagger$\\
\bottomrule
\end{tabular}
\end{center} 
{\scriptsize \dag indicates cases where Triton analyzed partial inputs from dataset.}
\caption{Edge coverage comparison of 5 taint-guided fuzzers for 24 hour time budget}
\label{edge-coverage}
\end{table}

In this section, we compare the performance of all tools when applied to one of the most important security applications, taint-guided fuzzing. 

\medskip
\noindent
\textbf{Fuzzer Backend Implementation.}
Dynamic taint analysis has been used as the front end by many fuzzers to identify hot bytes for guiding further mutations.~\cite{taintscope}~\cite{borg}~\cite{dowser}~\cite{vuzzer}~\cite{angora}. Although these fuzzers leverage dynamic taint analysis tools to find hot bytes, each of them applies different algorithms to mutate hot bytes. For example, Vuzzer~\cite{vuzzer} copies the magic number extracted from binary directly to the locations of hot bytes; and Angora~\cite{angora} implements gradient descent along with other strategies to mutate these hot byte locations. 
To eliminate the effects of different searching strategies and different execution backends of fuzzing modules, we build a simple and efficient fuzzer backend in C (shown in Appendix~\ref{app:fuzzing}). For each dynamic taint analysis front end, we use a common backend that generates new inputs based on mutations of the hot bytes. 
Therefore, we have a total of four fuzzer prototypes using the same mutation algorithm as shown in Algorithm~\ref{alg1}. The metric we use for comparison is the edge coverage achieved from the different front ends. 

\medskip
\noindent
\textbf{Edge Coverage Comparison.}
We run each fuzzer prototype for 24 hours on the 6 programs shown in Table~\ref{edge-coverage}. We have discussed the evaluation of hot byte accuracy for these programs in~\Cref{hot_byte_eval}. Each fuzzer is given the same initial seed corpus for each program and assigned a single CPU. Each dynamic taint analysis front-end works with the back-end fuzzer to guide the mutation. Since Triton incurs a significantly large runtime overhead greater the our evaluation time, Triton has only analyzed partial inputs in 24 hours. The results are shown in Table~\ref{edge-coverage}. \tool achieves the highest edge coverage for all 6 program evaluated. On average, \tool reaches $61\%$ more edge coverage than the second best analysis front end Libdft. Triton is the worst analysis front end due to its extremely large runtime overhead and lowest hot byte accuracy. DFSan front end could only run on two programs and achieves intermediate results. To summarize, the taint-guided fuzzing results further validate that \tool obtains the most accurate hot bytes among the four taint analysis tools in real-world applications. The consistency of taint-guided fuzzing results and hot byte accuracy evaluation (\Cref{hot_byte_eval}) demonstrates the efficiency and effectiveness of \tool.

\begin{longfbox}
\textbf{Result 5:} \tool achieves 61\% more edge coverage than other dynamic taint analysis tools for taint-guided fuzzing, demonstrating that the taint information obtained from \tool is more effective.
\end{longfbox}

\begin{table*}[ht!]
\centering
\footnotesize
\begin{tabular}{l|ccccc|ccccc}
\toprule
\multirow{2}{*}{\bf Program} & \multicolumn{5}{c}{\bf Hot Byte Accuracy} & \multicolumn{5}{c}{\bf Hot Byte FPR} \\
& {\bf NN} & \bf Logistic & \bf {SVM(linear)} & \bf SVM(poly) & \bf SVM(rbf) & \bf NN & \bf Logistic &\bf SVM(linear) & \bf SVM(poly) & \bf SVM(rbf)\\
\midrule
readelf & 81\% & 38.6\% & 23.8\% & 47.4\%& 27.4\% & 1.6\% & 4.2\% & 5.2\% &3.6\%&4.9\% \\
harfbuzz & 86\% &19.8\% &26.3\% & 56.7\% &28.7\%  & 0.8\% & 4.8\% & 4.4\%& 2.6\% & 4.2\% \\
mupdf & 80\% & 13\% & 14\% &  14.8\%& 17.4\% & 2\% & 5\% & 5\%& 4.9\%& 4.8\% \\
libxml & 65\% & 34.8\% &47\% & 42.3\%& 5.7\% &  2.3\% & 4.3\%& 3.5\% &3.9\%&6.3\%\\
libjpeg &  29\% & 7.3\% & 5.9\% &7\%& 1.4\% &  3.9\% & 7.3\% & 5.1\%& 5\% &  5\%\\
zlib & 66\% & 44\% &7.1\% & 15\%& 10\% & 1.8\% & 3\% &4.8\%& 4.4\% & 4.7\%\\
\bottomrule
\end{tabular}
\caption{\tool performance on different ML models. The neural network model achieves on average $68\%$ hot byte accuracy and $2.07\%$ FPR, the best among five machine learning models.}
\label{tab:models}
\end{table*}

\subsection{How does \tool perform with different machine learning models other than neural networks?}

In this section, we compare other machine learning models (\eg logistic regression and support vector machines (SVM)) against neural network for the implementation of \tool on the same set of real world programs. We also use the hot byte accuracy and FPR as mentioned in Section~\ref{hot_byte_eval} to evaluate the performance of different machine learning models.

\noindent
\textbf{Logistic Regression:} We implement the logistic regression using the same NN architecture, but without any non-linear activation in hidden layers. 
We use sigmoid as final layer and train with the same setting as mentioned in~\Cref{sec:exp_setup}. To extract the hot byte information, we perform the gradient analysis routine as mentioned in~\Cref{hot_byte_eval}.

\noindent
\textbf{SVM models:}
We implement the SVM models (linear, polynomial kernel, and RBF Gaussian kernel) using the scikit-learn library~\cite{sklearn}. Our dataset has a large number of output labels (\ie, up to 2K sink variables for each input). The runtime overhead is large for SVM model on such dataset because it uses a simple one-vs-one scheme. Therefore, we leverage the correlations between labels and encode the large number labels into a smaller and compact ones. To obtain the importance of each input feature, we compute the dot product of all weights associated with each input feature to the final model outputs. Bigger weights mean that the output is more sensitive to the change of the corresponding feature.

\noindent
\textbf{Results:}
Table~\ref{tab:models} shows the hot byte accuracy and false positive rate from the five different ML models. Neural network model achieves the best results on all 6 programs, on average $68\%$ hot byte accuracy. Among the four other machine learning models, SVM with polynomial kernel is the best model, achieving on average $30.5\%$ hot byte accuracy. Logistic model achieves the second-best results on programs \texttt{libjpeg} and \texttt{zlib}. SVM with polynomial kernel is the best-performed SVM model, achieving the second-best results on programs \texttt{harfbuzz} and \texttt{readelf}. SVM with linear kernel achieves the second-best result on program \texttt{libxml} and SVM with RBF Gaussian kernel achieves the second-best result on program \texttt{mupdf}. The reason for neural network's superior performance is that neural network has a large model capacity such that it fits diverse datasets well. Moreover, unlike SVM, neural network can naturally support datasets with a large number of labels. 

\begin{longfbox}
\textbf{Result 6:} The neural network model achieves on average $68\%$ hot byte accuracy and $2.07\%$ FPR, the best among five machine learning models.
\end{longfbox}

\subsection{What kinds of flows are missed by \tool? How does the training data quality affect such information loss and how to mitigate this loss?}
\label{flowloss}

\begin{figure}[t!]
\centering
\includegraphics[scale=0.35]{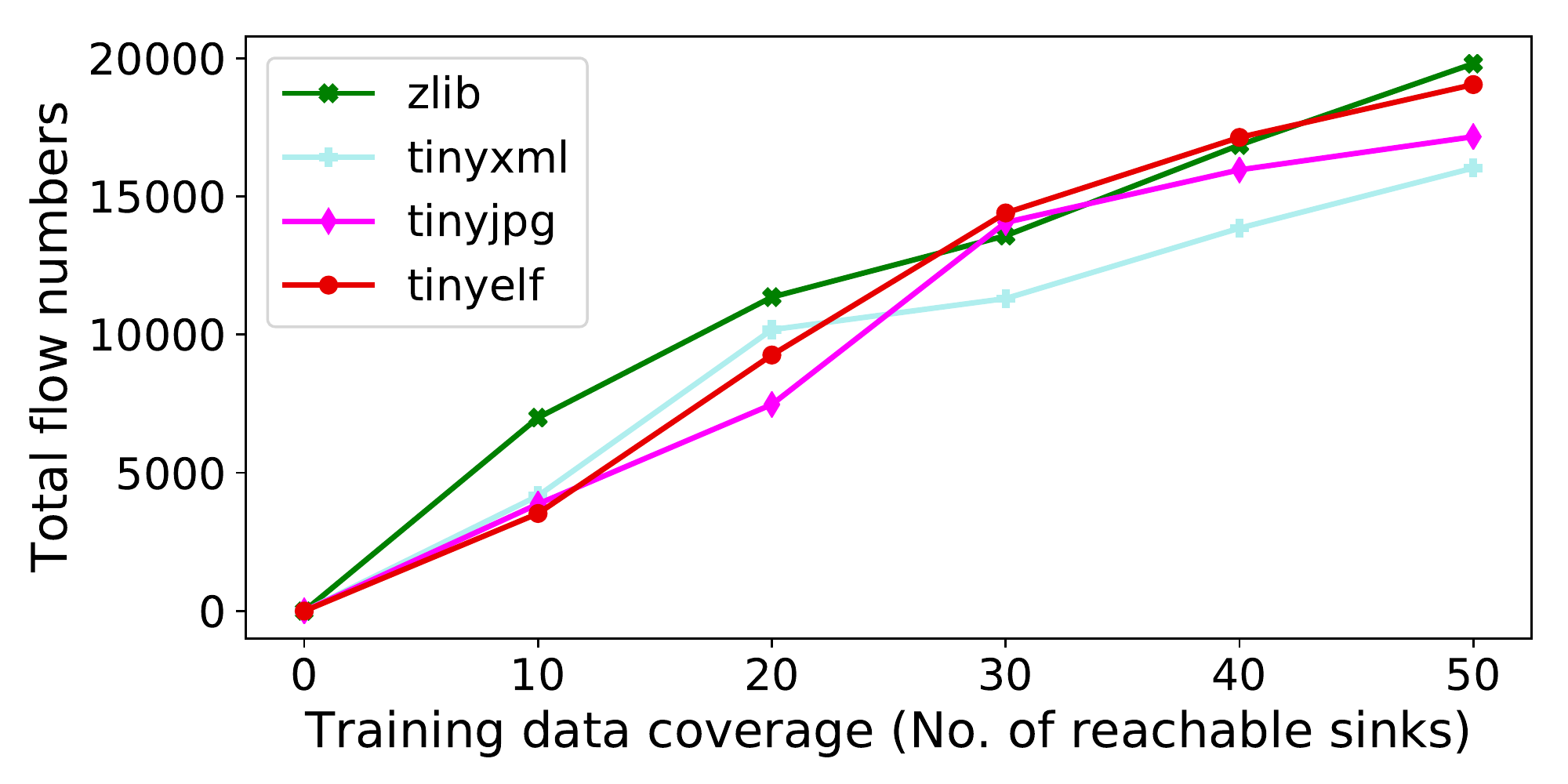}
\caption{Total number of flows detected by \tool when training dataset coverage increases. Training data with higher coverage can increase the flow coverage of \tool.}
\label{fig:info_loss}
\end{figure}

\noindent
\textbf{Flow Definition:} Before describing our techniques to measure the information loss, we formally define the flows. The goal of DTA is to detect flows based on dynamic execution. Therefore, we collect a ground truth dataset that contains the total number of flows based on \emph{unseen} test inputs. We define one flow as a tuple of (input value, source, sink), and the total flows are collected from all test inputs. To measure the information loss, we evaluate how many flows \tool and dynamic taint analysis tools can detect out of the total flows from the ground truth.

\medskip
\noindent
\textbf{Flow Dataset:}
Since we need the ground truth of exact number of total flows as baseline, we choose four programs (including three tiny programs~\cite{tinyxml,tinyelf,tinyjpg}, and one small real world program) where such information can be obtained reliably through static analysis and manual inspection. For each program, we choose 50 sink variables in conditional predicates, such as file magic bytes, field values and offsets. We obtain these sink values through light-weight instrumentation and normalize them into binary data as mentioned in~\Cref{sec:exp_setup}. These sink variables are commonly used to perform conditional checking that determines program behaviors. If a particular input can reach 20 sinks, we count 20 flows for that input. To collect the dataset, we randomly flip bytes of a specified input using a simple fuzzer, generating 6K inputs which cover all the 50 sink variables. Then we split the dataset into 5K training inputs and 1K testing inputs. Both training and testing datasets can cover all 50 sink variables. We achieve on average $99\%$ testing accuracy. After training the neural program, we perform the gradient analysis as mentioned in~\Cref{hot_byte_eval} to reason if a sink variable is tainted or not. If the gradient value for a sink variable is greater than a specified threshold, it is considered as tainted.

\medskip
\noindent
\textbf{Information Loss:}
All dynamic taint analysis tools suffer from information loss, since they cannot track all flows in a program. The information loss of \tool can be categorized into two classes. 
\begin{itemize}
	\item Coverage of training dataset. When the training data do not cover all the sink variables that appear in the testing data, there could be information loss on \tool.
	
	\item Inaccuracy of machine learning model. No model is $100\%$ accurate on unseen testing data. The information loss happens when the neural network model makes wrong predictions to unseen testing data.
\end{itemize}

To investigate the information loss caused by training data coverage (\ie, the number of sink variables covered by training inputs), we downsample the 5K training inputs into five subsets with different coverage threshold. Each subset covers a different number of sink variables from 10, 20, 30, 40, to 50. Then we train \tool with each subset and evaluate the total number of flows detected on the 1K \emph{unseen} test inputs. The result is shown in Fig~\ref{fig:info_loss}. The total number of flows detected by \tool increases as the training data coverage increases. When training data covers all the sinks, \tool can detect the highest number of flows in the unseen testing dataset.

\label{subsec:infoloss}
\begin{table}
\scriptsize
\centering
\begin{tabular}{lccccc}
\toprule
\multirow{2}{*}{\bf Program} & {\bf Total Flows} & \multirow{2}{*}{\bf \tool} & \multirow{2}{*}{\bf Libdft} & \multirow{2}{*}{\bf Triton} & \multirow{2}{*}{\bf DFSan} \\
& {\bf (Ground Truth)} &  &  &  &  \\
\midrule
TinyELF & 19,464 & 19,046 & 7,227 & 18,048 & 5,120 \\
TinyJPG &  17,188 & 17,160 & 11,439 & 17,184 & 15,510 \\
TinyXML & 17,036 & 16,691 & 15,233 & 7,671 & 14,720 \\
Zlib & 19,957 & 19,804 & 18,043 & 14,743& 14,322\\
\bottomrule
\end{tabular}
\caption{Comparison of information flow losses of different taint tracking tools on three tiny programs and one real world program. We use static analysis and manual examination to estimate the number of total flows.}
\label{tab:info_loss}
\end{table}

Furthermore, we evaluate the information loss caused by the inaccuracy of the neural network model. Specifically, we compare the number of flows detected by \tool against three state-of-the-art DTA tools on the four programs. We obtain the ground truth (\ie, total number of flows) from all the testing inputs.
The result is shown in Table~\ref{tab:info_loss}. On average, \tool detects $98.7\%$ flows, the highest among all tools. Triton, Libdft, and DFSan detect on average $78\%$, $70.9\%$ and $69\%$ flows, respectively. The $1.3\%$ loss of \tool is due to model inaccuracy.

\medskip
\noindent
\textbf{Advantages of \tool:}
The reason for \tool's superior performance is that \tool can detect information flows passed through some complex code such as external library calls strcmp() and strncmp(). It is hard for DTA tools to propagate the taint tags through the library calls. Moreover, \tool also detects some implicit control flow dependency which is not supported by common DTA tools. We will cover more details of these cases in~\Cref{undertaint_case_study} and Appendix~\ref{app:case_study}.
\tool achieves the best result on 3 programs (\texttt{TinyELF}, \texttt{TinyXML}, and \texttt{Zlib}) among all four tools and the second best on \texttt{TinyJPG}.
For the program \texttt{TinyJPG}, Triton finds slightly more flows than \tool. \texttt{TinyJPG} does not contain any external library calls or implicit control dependency in selected taint sinks, while \tool could make some minor mistakes when approximating sink variables in unseen inputs. Lifdft performs the worst on the program \texttt{TinyELF}, because most functions in \texttt{TinyELF} pass parameters through a float point instruction MOVSD which is not supported by Libdft. Compared to other tools, DFSan finds the least number of flows. The results show that information loss is common for all tested tools.



\medskip
\noindent
\textbf{How to improve the coverage of training data for \tool?} To mitigate the information loss, we can add more training data to reach more sink variables. Using a fuzzer to generate data with coverage guidance is more helpful than only randomly flipping the input without the guidance. In addition, we can use existing techniques such as symbolic execution to generate training data with high quality. Finding a new path in any form of DTA is the hard problem. Once we find one path between source and sink, we can generate many more paths via input mutation. Though training the neural program requires at least one path between a source and a sink, we can achieve lower false positive rate and higher accuracy than rule-based traditional DTA tools.

\begin{longfbox}
\textbf{Result 7:} Training data with higher coverage increases the flow coverage of \tool. On average, \tool detects $98.7\%$ flows which is $20\%$ more than the second-best tool Triton.
\end{longfbox}

\section{Undertaint Case Study}
\label{undertaint_case_study}
In this section, we present a case to explain why \tool is more accurate than traditional dynamic taint analysis. Specifically, \tool tracks implicit information flows and avoids under taint in real world programs.
More examples can be found in Appendix~\ref{app:case_study}.

\medskip
\noindent
\textbf{Example: Implicit Information Flows.}
Most DTA tools ignore implicit information flows (\ie, implicit control dependency and complex external library calls) and only support explicit information flow in data-dependency form. The reason is that supporting implicit information flows could cause high runtime overhead and false positive rate~\cite{libdft}. Lack of support to these implicit information flows often result in under-taint issue on some real world programs. We discuss the implicit control dependency in a popular tiny program \texttt{TinyXML}~\cite{tinyxml}. As shown in List~\ref{lst:implicit_control}, \texttt{p} is a input buffer which stores program input as taint source. At line 16, \texttt{ele-\string>ClosingType()} is the sink variable determining the program branching behavior. At line 6, \texttt{ele-\string>ClosingType()} can be modified to a constant value when there is a special character in the input buffer. Since \texttt{ele-\string>ClosingType()} is control-dependent but not data-dependent on taint source input buffer \texttt{p}, DTA tools fail to track the flows to such sink variables. In the second example, we consider such implicit control dependency in complex external function calls. Many sink variables are the return values of complex external function calls such as strcmp() and strncmp(). The return values are state variables which are control-dependent on the taint source, but not explicit data dependent. Therefore, DTA tools would lose taint information for such sink variables.

\lstset{basicstyle=\footnotesize\ttfamily,breaklines=true}
\begin{lstlisting}[caption={Implicit control dependency in \bf\texttt{TinyXML}}, captionpos=b, label={lst:implicit_control}, frame=None, belowcaptionskip=.01cm, belowskip=1pt]
// tinyxml2/tinyxml2.cpp:1066
while(p){
  ...
  if( *p == '/' ) 
  {
    /* implicit control flow dependency*/
    ele->ClosingType() = CLOSING;
    ++p;
  }
}

char* XMLNode::ParseDeep(...)
{
  ...
  /* under-taint */
  if(ele->ClosingType() == XMLElement::CLOSING)
  {
    ...
  }
  ...
}
\end{lstlisting}
\textbf{\tool Solution.}
\tool avoids this problem by directly learning the mapping from taint sources to any taint sink variables (including both data-dependency and control-dependency variables). Compared to traditional dynamic taint analysis tools, \tool has the advantage of generalizing to various real-world programs.

\section{Related Work}


Recently, many works~\cite{BhatiaS16,Pu2016,GravesWD14,pmlr-v37-piech15,Reed2015NeuralP} have used machine learning for different program analysis tasks such as program synthesis~\cite{Parisotto2016NeuroSymbolicPS}, vulnerability detection~\cite{Hovsepyan_2012,Pang2015,Mou2016,LiZXO0WDZ18,Huo2016,Lam2015,Choi_2017}, program repair~\cite{neural_repair,deepfix}, fuzzing~\cite{rajpal2017not,godefroid2017learn,Nichols2017FasterFR,smart_seed,gan_fuzz,neuzz}, and symbolic execution~\cite{shen2018neuro}.

Dynamic taint analysis~\cite{Newsome05dynamictaint}~\cite{Bosman2011MinemuTW}~\cite{Yin07panorama}~\cite{libdft}~\cite{MingStraightTaint} executes programs with concrete inputs to perform the analysis. However, it incurs large overhead and suffers from overtaint and undertaint issues. To address these issues, TaintInduce~\cite{taintinduce} proposes to learn platform-specific taint propagation rules from (input, output) pairs of instructions. Their approach learns propagation rules based on a template, and uses an algorithm to reduce the task to learning different input sets and pre-conditions for propagating the taint tags. TaintInduce increases the accuracy for individual propagation rules, but they still suffer from accumulated errors and large overhead due to propagation-based design.
On the contrary, \tool uses machine learning technique to track end-to-end information flow. We use light-weight instrumentation to build neural program embeddings, and directly analyze the flow of information captured by the neural network models. Our technique minimizes end-to-end information flow tracking errors and significantly reduces runtime overhead.



\section{conclusion}

We present a novel approach \tool to perform taint analysis using neural program embeddings. Our neural program learns the information flow directly from taint sources to taint sinks. We use saliency maps to analyze the information flow in the neural programs. To evaluate the accuracy, overhead, and application utility of \tool, we compare against three state-of-the-art dynamic taint analysis tools. The results show that \tool achieves on average $10\%$ increase in accuracy and 40 times less runtime overhead over the second best dynamic taint analysis tool Libdft. \tool can also successfully track the information from source to sink in exploits. We further evaluate \tool through a popular taint application--fuzzing. The taint-guided fuzzing results demonstrate that \tool can achieve on average $61$\% more edge coverage than state-of-the-art dynamic taint analysis tools.
\section*{acknowledgement}
We thank Mingshen Sun, our shepherd Mathias Payer and the anonymous reviewers for their constructive and valuable feedback. This work is sponsored in part by NSF grants CNS-18-42456, CNS-18-01426, CNS-16-17670, CNS-16-18771, CCF-16-19123, CCF-18-22965, CNS-19-46068; ONR grant N00014-17-1-2010; an ARL Young Investigator (YIP) award; a NSF CAREER award; a Google Faculty Fellowship; and a Capital One Research Grant. Any opinions, findings, conclusions, or recommendations expressed herein are those of the authors, and do not necessarily reflect those of the US Government, ONR, ARL, NSF, Google, or Capital One.
\bibliographystyle{abbrv}
\bibliography{paper.bib}

\normalsize
\appendix
\subsection{NN Architecture}
\label{app:nn_arch}
\begin{figure}[H]
\centering
\includegraphics[scale=0.6]{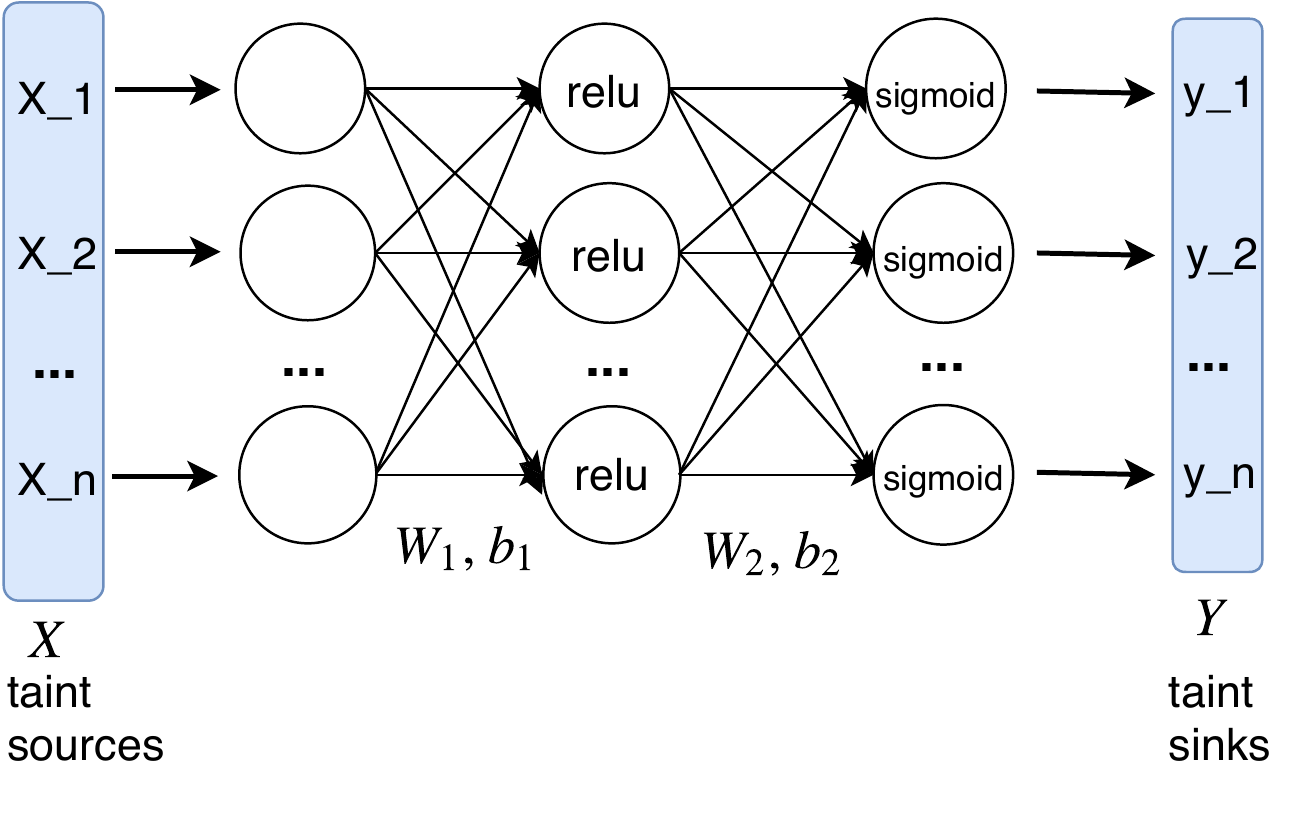}
\caption{The Neural Network architecture we use to generate dynamic program embeddings.}
\label{fig:architecture}
\end{figure}

We use the neural network architecture shown in Figure~\ref{fig:architecture} to learn neural program embeddings. The taint sources are \nn inputs and taint sinks are \nn outputs. The network has one hidden layer with ReLU activations and one output layer with sigmoid activations. 

\subsection{Saliency Map for ELF File Format}
\label{app:saliency}
\begin{figure}[H]
\centering
\includegraphics[scale=0.28]{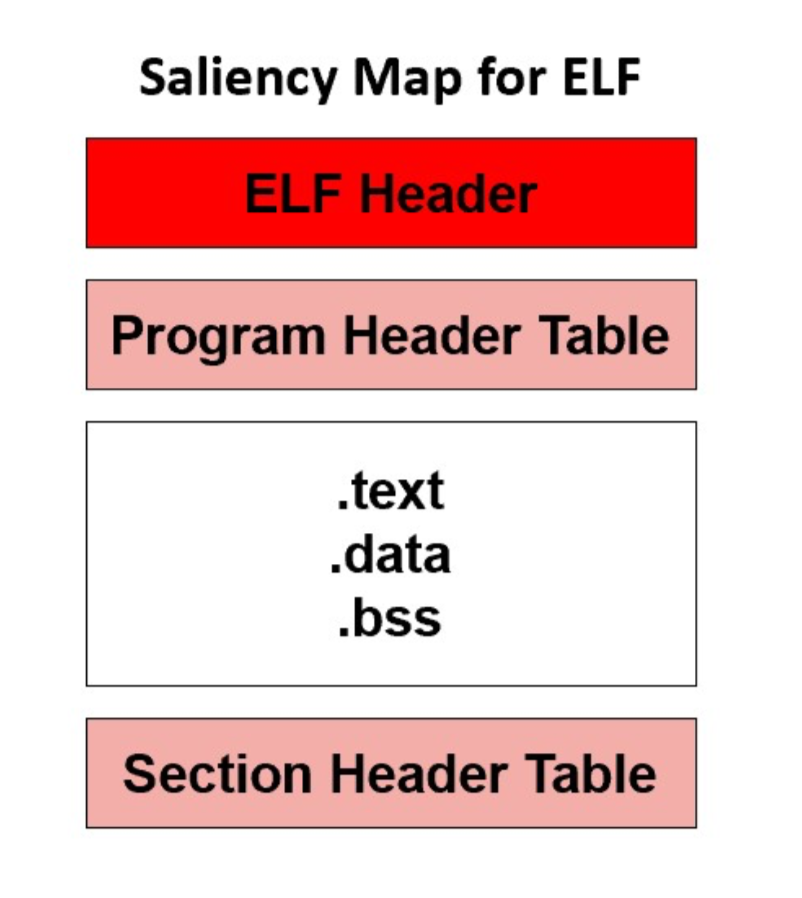}
\caption{Saliency Map of user input on program {\bf\texttt{readelf}}. The darkness of the color corresponds to the influence of the region's bytes on the taint sinks. The ELF Header field has the largest saliency values.}
\label{fig:saliency_map}
\end{figure}

The ELF file format can be broken down into four main regions, as shown in Fig~\ref{fig:saliency_map}. Three of them are header information (ELF, Program Header Table, and Section Header Table) and the other one is for non-header information (\eg .text, .data, .bss). For a common ELF file parser (\texttt{readelf}), the main parsing logic focuses on these shaded header regions and typically ignores the non-header information. Therefore the hot bytes should locate at the these header regions.

\subsection{Fuzzing Algorithm}
\label{app:fuzzing}
Our fuzzing algorithm is shown in Algorithm~\ref{alg1}. The algorithm takes in hot bytes locations from taint tools and perform deterministic mutations on these hot bytes. Random mutations are discarded to ensure that the fuzzer performance is only affected by the quality of hot bytes identified by the different dynamic taint analysis front ends. 
\begin{algorithm}[H]
\caption{Our mutation algorithm for taint-guided fuzzing that focuses on influential bytes.} 
\label{alg1} 
\lstset{basicstyle=\ttfamily\footnotesize, breaklines=true}
\begin{tabular}{|lp{2.6in}|}\hline
\textbf{Input}:
    & \textit{seed} $\leftarrow$ initial seed \\
    & \textit{iter} $\leftarrow$ number of iterations \\
    & \textit{hot\_bytes} $\leftarrow$ hot bytes from taint tools \\
\hline
\end{tabular}
\begin{algorithmic}[1] 
\FOR{$i = 1$ to $iter$}
        \STATE $locations \gets top(hot\_bytes,(2^i))$
        \FOR{$m = 1$ to $255$}
            \FOR{$loc \in locations$}
                \STATE $v \gets seed[loc] +  m$
                \STATE $v \gets clip(v, 0, 255)$
            \ENDFOR
            \STATE $gen\_mutate(seed, loc, v)$
            \FOR{$loc \in locations$}
                \STATE $v \gets seed[loc] -  m$
                \STATE $v \gets clip(v, 0, 255)$
            \ENDFOR
            \STATE $gen\_mutate(seed, loc, v)$
        \ENDFOR
\ENDFOR
\end{algorithmic}
\end{algorithm}

\subsection{Case Study}
\label{app:case_study}
In this section, we present case studies to explain why \tool is more accurate and has lower runtime overhead than traditional dynamic taint analysis tools.

\subsubsection{Undertaint}
When a taint analysis tool fails to track all the taint labels for a specified variable, it is considered as under-taint. Under-taint is a common issue for dynamic taint analysis. Since the taint propagation rules in dynamic taint analysis tools are neither sound nor complete, some taint labels are easily missed during the analysis process~\cite{taintinduce}.

\label{app:undertaint}
\medskip
\noindent
\textbf{Example: Pointer Taint.}
In the code example~\ref{lst:under_taint}, we demonstrate the classic pointer taint dilemma in a popular XML parser library \texttt{libxml}, where we track the taint flow from program input to \texttt{NXT(len)} (line $7$ and $8$). From the prior execution context, variables \texttt{ctxt->cur} and \texttt{len} are all affected by program input taint source and therefore carry the taint label of the taint source. The example shows a function, \texttt{xmlXPathComPathExpr()}, that is frequently used by the library to parse the path expression of a XML element. It uses \texttt{len} as the index to check the one-character operator of a path expression at line $7$ and $8$, through the byte reading macro \texttt{NXT(val)} defined in line $2$. After taking \texttt{len} as the offset to the current pointer location \texttt{ctxt->cur}, \texttt{NXT()} returns the byte located at the address \texttt{ctxt->cur + len}. The propagation rules state that the byte memory is only affected by a single byte read from the memory content \texttt{NXT(val)}, not by the base address \texttt{ctxt->cur} and \texttt{len}. However, these addresses determine the byte memory content,
which are missed due to pointer under-taint. In practice, the taint flow from pointer to memory content is intentionally ignored by most taint analysis tools as handling them could easily cause many false positives and even a taint explosion.\cite{pointer_taint} In contrast, \tool can capture such information flow from pointer to program behavior. Specifically, \tool models the function mapping from program input to branch variables at line $7$ and $8$. Then, by learning the differences among program behavior triggered by various input samples at line $7$ and $8$, \tool can infer input bytes that reach line $10$. The advantage of our method over traditional taint analysis is that it is based on the knowledge learned from a summary of runtime program semantics rather than fixed taint propagation rules. 

\lstset{basicstyle=\footnotesize\ttfamily,breaklines=true}
\begin{lstlisting}[caption={Pointer under-taint in \bf\texttt{libxml}}, captionpos=b, label={lst:under_taint}, frame=None, belowcaptionskip=.01cm, belowskip=1pt]
// libxml2-2.9.7/xpath.c:10736
#define NXT(val) ctxt->cur[(val)]

static void xmlXPathCompPathExpr(...)
{
  ...
  if((NXT(len) == '<') || (NXT(len) == '>') 
  || (NXT(len) == '='))
  {
    lc = 1;
    break;
  }
  ...
}
\end{lstlisting}

\medskip
\noindent
\textbf{Example: Incomplete Taint Source.}
Incomplete taint source identification can also cause severe under-taint issue in real-world applications. Dynamic taint analysis tools identify the taint source by installing some hooks to system calls \texttt{open()}, \texttt{read()}, \texttt{mmap()} and setting taint marks on corresponding memory/registers. These predefined interception procedures usually rely on developers' experience to cover some common cases. However, real-world applications have diverse and complex IO procedures. The simple system call hooks in dynamic taint analysis tools may fail to fully capture all the taint sources and lose track of taint information at the beginning of program execution. For example, a popular XML parser library \texttt{libxml} supports uses compressed IO interface \texttt{gzread()} by default for both compressed and uncompressed input, which a lot of tools are not aware of. In this case, a state-of-the-art dynamic taint tool libdft would lose track of partial taint source and result in a severe under-taint problem. To make the matter worse, many different under-taint reasons can co-occur at different parts of the program.

\medskip
\noindent
\textbf{\tool Solution.}
\tool avoids this problem by not relying on any human engineered system call hook procedures. It directly uses a neural network model to learn the mapping between taint sources and taint sink variables. Compared to traditional dynamic taint analysis tools, \tool has the advantage of generalizing to various real-world programs.

\subsubsection{Overtaint}
\label{app:over_taint_text}
Over-taint occurs when dynamic taint analysis marks irrelevant taint labels on specified variables. Dynamic taint analysis tools define over-approximated taint propagation rules for certain types of instructions, making it hard to track the precise taint flow. Since the taint labels are propagated at the instruction level and often ignore the semantic of the program context, the over-taint issue is inevitable. Furthermore, in a real-world program, a single over-taint label could instantly propagate to generate many over-taint labels during a repeating operation (\eg a loop or recursive function call). This over-taint issue pollutes further execution and analysis results.

\lstset{basicstyle=\footnotesize\ttfamily,breaklines=true, xleftmargin=0.65cm, numbers=left}
\begin{lstlisting}[caption={Over-taint propagation in \bf\texttt{zlib}}, captionpos=b, label={lst:over_taint}]
// zlib-1.2.11/inffast.c:50
void ZLIB_INTERNAL inflate_fast(strm, start)
{
  ...
  for(...)
  {
    hold += (unsigned long)(*in++) << bits;
    bits += 8;
    hold += (unsigned long)(*in++) << bits;
    bits += 8;
    ...
    /* decoding match distance */
    if(dodist)
    {
      dist=(unsigned)hold & ((1U << op) - 1);
      if(dist > dmax) {...}
    }
    ...
    hold >>= (bits + 16); // over-taint
  }
  ...
}
\end{lstlisting}
\vspace{0.6cm}

\medskip
\noindent
\textbf{Example. Error Accumulation.}
As shown in code~\ref{lst:over_taint}, \texttt{inflate\_fast()} is a decoding function in a \texttt{zlib} decompression procedure, where we track the taint flow from input buffer \texttt{in} (line $7$ and $9$) to the distance variable \texttt{dist} (line $16$). The function reads compressed data from input buffer \texttt{in} and decodes the distance variable \texttt{dist} for further deflation. Going through the for loop, every time two bytes are read from \texttt{in} buffer (lines $7$-$10$) and stored into the bit accumulator \texttt{hold}, then the content in \texttt{hold} are dropped at line $19$. After a few iterations, the condition at line $13$ is satisfied and the function starts to decode the match distance variable \texttt{dist}. So \texttt{dist} is only affected by the two new bytes in \texttt{hold} that are read from the current round. The over-taint occurs at line $19$ when \texttt{hold} drops bits from the previous round through a shift operation, but the taint rules did not drop the labels. Since no taint propagation rule can be general enough for different program semantic, dynamic taint analysis tools use a conservative taint rule to copy all the taint labels from source (\texttt{hold}) to destination (\texttt{hold}) for shift instructions. For every round in the loop, \texttt{hold} obtains two byte taint labels while still keeping the old taint labels from the previous round. The total taint labels of \texttt{hold} were accumulated through the loop, then spread to \texttt{dist} that determines the conditional program behavior at line $16$. The over-taint propagation issue stems from the inaccurate taint propagation rules pre-defined by experts. It is impossible to design fixed taint propagation rules to accurately handle all the real-world cases.

\medskip
\noindent
\textbf{\tool Solution.}
\tool directly models the mapping from program input to conditional program behaviors at line $16$. As long as there exist samples that cover different branches at line $16$, \tool can easily infer the two critical bytes of program input that affect \texttt{dist}.

\subsubsection{Overhead}
\label{app:overhead}
Traditional dynamic taint analysis tools track the taint labels by instrumenting the execution of the program. Specifically, for every executed instruction, the taint analysis tool calls a corresponding taint propagation handler to process the taint labels associated with the operation. Most instructions trigger calls to these handlers to track the taint flow with very few exceptions (\eg push, pop and ret), leading to a large runtime overhead. The runtime overhead of taint propagation is also affected by the granularity of the taint tag. Marking taint tags for each input on the coarse binary taint tag granularity already incurs significant overhead, not to mention the overhead to mark each input offset. A complex dynamic taint analysis task such as fuzzing requires more fine-grained taint labels which represents different offsets of user input. Tracking numerical taint labels incurs more runtime overhead than simple binary taint labels due to the fact that a basic union operation over two tainted source tags with respectively $m$ and $n$ offsets is O(mn) time complexity, rather than O(1) on two binary taint labels. As a result, adopting fine-grained taint labels drastically increases the runtime overhead of taint propagation handlers.      

\medskip
\noindent
\textbf{Example.}
Real-world programs can have many time-critical operations that are frequently executed. If every instruction of these time-critical procedures needs to call additional taint propagation handler, then the runtime overhead would be significantly large.
As shown in code ~\ref{lst:run_time}, \texttt{decode\_mcu()} is a common function used in JPEG parser library \texttt{libjpeg}, where we track taint flow from program input to DC coefficients \texttt{block}. The function uses a for loop to decode Huffman-compressed coefficients, by repeatedly calling a function macro \texttt{HUFF\_DECODE} at line 17. Since \texttt{HUFF\_DECODE} is a time-critical operation, it is implemented as inline-macro for better performance. However, most operations in \texttt{HUFF\_DECODE} involve taint propagation. Starting from line $5$, the macro reads a byte \texttt{c} from input buffer, then performs a binary OR between \texttt{get\_buffer} and \texttt{c} at line 6. So \texttt{get\_buffer} takes the union of taint labels for \texttt{c} and itself. At line $7$, the global variable \texttt{get\_buffer} is used to perform huffman decoding which requires to propagate the taint label from \texttt{get\_buffer} to \texttt{result}. Repeated calls to the macro \texttt{HUFF\_DECODE} all involve taint label propagation, causing extreme runtime overhead. These taint labels finally propagates into result variable \texttt{s} at line 20, which would be intensively used during later decoding procedure and introduce even more runtime overhead. Our experiments show that a state-of-the-art dynamic taint analysis tool has more than 10,000X runtime overhead than normal execution on libjepg.

\medskip
\noindent
\textbf{\tool Solution.}
As mentioned in last section, \tool is a black-box analysis that does not need to track every instruction, enabling it to avoid large runtime overheads.

\lstset{basicstyle=\footnotesize\ttfamily ,breaklines=true ,xleftmargin=0.65cm, numbers=left}
\begin{lstlisting}[caption={Extreme runtime overhead in \textbf{\texttt{libjpeg}}. In Line 17, macro \texttt{HUFF\_DECODE} is repeatedly called, which involves expensive taint label propagation, causing extreme runtime overhead.}, captionpos=b, label={lst:run_time}]
// jpeg-9c/jdhuff.c:1197
#define HUFF_DECODE(result, ...) \
{
  ...
  c = read_byte(); \
  get_buffer = get_buffer | c; \
  result = huff_decode(get_buffer, ...); \
  ...
}

bool decode_mcu(j_decompress_ptr cinfo, ...)
{
  ...
  for(...)
  {
    ...
    HUFF_DECODE(s, br_state, htbl, ...);
    ...
    /* Output the DC coefficient */
    (*block)[0] = (JCOEF) s; 
    ...
  }
  ...
}
\end{lstlisting}

\end{document}